\documentclass[letterpaper,journal]{IEEEtran}
\usepackage{amsmath,amsfonts}
\usepackage{algorithmic}
\usepackage{algorithm}
\usepackage{array}
\usepackage[caption=false,font=normalsize,labelfont=sf,textfont=sf]{subfig}
\usepackage{textcomp}
\usepackage{stfloats}
\usepackage{url}
\usepackage{verbatim}
\usepackage{graphicx}
\usepackage{subfig} 
\usepackage{cite}
\usepackage{dcolumn}
\usepackage{booktabs}
\usepackage{pifont}
\usepackage{multirow}
\usepackage{graphicx}
\usepackage{subcaption} 
\begin{document}

\title{Bridging Speech Emotion Recognition and Personality: Dataset and Temporal Interaction Condition Network}

\author{Yuan Gao,~\IEEEmembership{Student~Member,~IEEE}, Hao Shi,~{Member,~IEEE}, Yahui Fu,~{Member,~IEEE},
\\ Chenhui Chu,~{Member,~IEEE}, and Tatsuya Kawahara,~\IEEEmembership{Fellow,~IEEE,}
\thanks{Yuan Gao, Hao Shi, Yahui Fu, Chenhui Chu, and Tatsuya Kawahara are with the Department of Intelligence Science and Technology, School of Informatics, Kyoto University, Kyoto, Japan. 
E-mail: gao@sap.ist.i.kyoto-u.ac.jp}}

\markboth{Journal of \LaTeX\ Class Files,~Vol.~14, No.~8, August~2021}%
{Shell \MakeLowercase{\textit{et al.}}: A Sample Article Using IEEEtran.cls for IEEE Journals}


\maketitle

\begin{abstract}
This study investigates the interaction between personality traits and emotion expression, exploring how personality information can improve speech emotion recognition (SER). We collect the personality annotation for the IEMOCAP dataset, making it the first speech dataset that contains both emotion and personality annotations (PA-IEMOCAP), and enabling direct integration of personality traits into SER. Statistical analysis on this dataset identified significant correlations between personality traits and emotional expressions. To extract finegrained personality features, we propose a temporal interaction condition network (TICN), in which personality features are integrated with HuBERT-based acoustic features for SER. Experiments show that incorporating ground-truth personality traits significantly enhances valence recognition, improving the concordance correlation coefficient (CCC) from 0.698 to 0.785 compared to the baseline without personality information. For practical applications in dialogue systems where personality information about the user is unavailable, we develop a front-end module of automatic personality recognition. Using these automatically predicted traits as inputs to our proposed TICN model, we achieve a CCC of 0.776 for valence recognition, representing an 11.17\% relative improvement over the baseline. These findings confirm the effectiveness of personality-aware SER and provide a solid foundation for further exploration in personality-aware speech processing applications.
\end{abstract}

\begin{IEEEkeywords}
Speech emotion recognition, Big Five personality traits, human computer interaction.
\end{IEEEkeywords}

\section{Introduction}
\IEEEPARstart{S}{peech} emotion recognition (SER) is widely known as a vital component of natural human–computer interaction \cite{cowie2001emotion,ramakrishnan2013speech}. Emotional information in speech reveals not only the immediate affective state of the speaker but also provides insight into how we think, feel, and behave \cite{vernon2006thinking}.
By enabling intelligent systems to detect and respond to the emotional state of the user, SER makes interactions more intuitive and improves the overall user experience.
Applications of SER range from empathetic speech assistants \cite{chatterjee2021real} and personalized systems for emotion-aware healthcare and automotive interfaces \cite{dhuheir2021emotion}, underscoring its significance within the broader field of affective computing.

Moreover, how we express and manage emotions is significantly influenced by our personality traits \cite{stemmler2010personality}. 
These traits shape cognitive processes, reactions, and communication styles, and thus play a key role in social interactions and emotional expression. 
Among various personality models \cite{cattell1992handbook, ashton2007empirical}, \cite{myers2010gifts}, the Big Five traits \cite{costa1999five,costa2008revised} is the most widely adopted framework in psychological research \cite{azucar2018predicting,simha2020big}. 
It defines five traits: openness (OP), conscientiousness (CO), extraversion (EX), agreeableness (AG), and neuroticism (NE). 
Each trait represents a distinct dimension of personality and offers a practical solution to quantify individual differences \cite{langston1997beliefs}. 
Previous studies have shown that these traits can substantially influence emotion perception, regulation, and expression \cite{deniz2011investigation, curtis2015relationship}. 
For instance, individuals high in EX tend to express their emotions more openly and with higher arousal, whereas those high in NE are often more sensitive to negative emotions and exhibit greater emotional instability. 

In this work, we annotate the well-known Interactive Emotional Dyadic Motion Capture (IEMOCAP) dataset \cite{busso2008iemocap} with Big Five traits. 
This personality-annotated IEMOCAP (PA-IEMOCAP) dataset, to our knowledge, is the first dataset of this type for speech, enables direct investigation of the interaction between personality and emotion expression\footnote{Data available at \url{https://github.com/Kyoto-University-Speech-and-Audio/PA-IEMOCAP}.} 
We first examine the correlation between each Big Five trait and the emotional dimensions of valence and arousal. Additionally, we incorporate a straightforward linear layer to project personality traits as feature embedding, which is integrated with acoustic features to improve the performance of SER. 

In practical applications, explicit personality information about users is unavailable \cite{wu2017implicit}. In this work, we explore effective models that incorporate automatically predicted personality traits for improving SER. Our approach begins by establishing a baseline for personality recognition (PR) at the utterance level. We then conduct multi-task learning experiments, jointly training PR and SER to explore their mutual influence within a shared feature extractor. Given that personality traits remain consistent throughout a conversation, we further provide a conversational-level PR baseline. 
These experiments reveal the potential of real dialogue systems in personalizing response according to the personality of users. 

Additionally, the predicted personality traits can serve as input of condition network, replacing ground-truth personality labels and thereby enable dialogue system better interpret and recognize emotional states about user. 
To better incorporate this predicted personality information for SER, we introduce a temporal interaction condition network (TICN). Although personality remains consistent over time, its impact on emotion expressions can vary across different speech segments. The proposed TICN is designed to capture this temporal variation, extracting finegrained personality features that enhance SER performance. Instead of combining features by concatenation, we incorporate cross attention mechanism \cite{vaswani2017attention}, allowing effective interaction between personality and acoustic features.

The main contributions of our work are summarized as:
\begin{itemize}
    \item \textbf{PA-IEMOCAP dataset:} We augment the widely-used IEMOCAP dataset with Big Five personality trait annotations, providing the first speech dataset that includes both emotion and personality labels.
    \item \textbf{Personality Recognition:} We conduct PR experiments on PA-IEMOCAP dataset at both the utterance and conversation levels, establishing baseline performance for PR.
    \item \textbf{Personality-Conditioned SER Models:} We propose TICN for incorporating annotated personality traits to enhance the SER model. We also evaluate whether leveraging predicted personality labels can improve SER performance when the personality information about the user is unavailable.
    \item \textbf{Multi-task Learning Frameworks:} We explore multi-task learning frameworks for jointly modeling SER and PR, allowing us to investigate the mutual influence between these two tasks.
\end{itemize}

\section{Related Works}
\subsection{Attribute Information for SER}
SER has attracted increasing attention in recent years \cite{khan2025joint}. Prior studies have shown that including additional speaker attributes such as gender \cite{li2019improved}, vocal characteristics \cite{triantafyllopoulos2021deep}, and linguistic content \cite{cai2021speech} can improve SER model performance. Researchers introduce multi-task learning to integrate these speaker attributes into SER models \cite{collobert2008unified, standley2020tasks}, \cite{zhang2021survey}. By simultaneously optimizing multiple related tasks, multi-task learning enables models to learn shared features and better utilize complementary information for emotion recognition \cite{zhang2017cross, sener2018multi}.
This approach has shown promising performance in SER \cite{sharma2022multi, latif2022multitask} benefiting from the extra information provided by emotion-related tasks. 

In multimodal emotion recognition, combining speech and text consistently outperforms speech-only systems, as linguistic information provides complementary semantic cues. Khan et al. \cite{khan2025memocmt} introduced a cross-modal transformer that integrates HuBERT-based speech features with BERT-based text features, achieving state-of-the-art results on benchmark datasets. These findings highlight the importance of jointly modeling acoustic and linguistic cues for emotion recognition. When explicit transcripts are unavailable, automatic speech recognition (ASR) can serve as an auxiliary task to implicitly capture linguistic information from speech. Cai et al. \cite{cai2021speech} proposed a multi-task learning framework combining ASR and SER with the wav2vec 2.0 feature extractor, where joint training effectively leveraged linguistic information and achieved state-of-the-art performance on IEMOCAP. Their ablation study further demonstrated that carefully weighting the ASR loss was critical to performance gains.
Additionally, speaker attributes have been widely studied: different speakers express emotions differently, affecting both acoustic and linguistic content for the same emotions. Thus, speaker-dependent SER systems usually outperform speaker-independent ones \cite{gat2022speaker}. As a broad category of speakers, gender information can also improve performance in speaker-independent SER \cite{li2019improved, pappagari2020x}.
More recently, Sharma et al. \cite{sharma2022multi} proposed a multi-task learning approach using pretrained self-supervised learning feature extractor. Their study showed further advantages of multi-task learning by integrating additional tasks like language classification and regression tasks related to pitch and energy. These additional tasks improved SER performance significantly across 25 datasets covering 13 languages and 7 emotion categories, demonstrating the effectiveness of multi-task learning in enhancing model generalization, especially in multilingual settings.
Despite the known relationship between personality traits and emotional expressions, personality remains an underexplored auxiliary attribute in SER.
\subsection{Joint Analysis of Personality and Emotion in Speech}
Personality traits and emotional states are closely connected, both influencing how people express themselves \cite{sagha2017effect,guidi2019analysis}. Understanding how these two aspects interact can lead to better performance in SER tasks. Recently, researchers have started exploring joint analysis frameworks to capture the interaction between personality and emotion, even though the lack of datasets annotated with both attributes remains a major challenge.
Zhang et al. \cite{zhang2019persemon} proposed PersEmoN, a deep multi-task learning framework for jointly analyzing personality and emotion. 
Because no publicly available corpus contained both emotion and personality annotations, they used two separate datasets (one labeled for emotion and another for personality) and incorporated an adversarial learning method to mitigate the mismatch of different datasets. 
Their results highlighted an inherent relationship between personality and emotion: personality shows stable impact on how emotions are expressed and managed. 
For example, they observed that EX corresponds to more intense emotions, while NE correlates with more frequent negative expression. 
Similarly, Li et al. \cite{li2023transfer} explored the personality-emotion connection in a transfer learning manner. 
They trained a wav2vec2-based model \cite{baevski2020wav2vec} for SER and then finetuned it for PR, finding that emotional features, particularly the arousal dimension, significantly improve PR performance. 
These studies underscore the effectiveness of incorporating personality information into SER. 

However, due to the lack of publicly available datasets that include annotations for both emotion and personality, previous works rely on separate data sources leading to substantial domain mismatch, and thus results in low generalizability of their findings.
\section{Personality Annotation for IEMOCAP dataset}
\label{sec3}
We conducted personality annotations for the widely-used IEMOCAP dataset to explore relationships between personality traits and emotional expressions.
The IEMOCAP dataset \cite{busso2008iemocap} is a benchmark dataset extensively used in emotion recognition and sentiment analysis research. It comprises approximately 12 hours of multimodal data collected from 10 professional actors (5 males and 5 females), including high-quality audio recordings, video, detailed facial motion capture data, and textual transcriptions.
Data collection involved five dyadic sessions, each including one male and one female actor performing scripted conversations and engaging in improvisational scenarios designed to elicit specific emotional expressions. The audio was captured using two microphones at a 48 kHz sampling rate and subsequently downsampled to 16 kHz to align with the common audio processing standard.
Below, we introduce key aspects of our annotation method, consistency evaluation, and correlation analysis.
\subsection{Labeling Procedure}
To elicit genuine emotional expressions and provide a more natural representation of affective behavior, the IEMOCAP dataset employs professional actors who perform diverse roles across conversations. We assume that the actors portrayed different characters in a manner closely approximating real-life behavior, and we therefore annotated the Big Five personality traits at the conversation level. These annotations reflected the enacted personalities in each conversation, rather than the actual personalities of the speakers, resulting in 302 personality profiles derived from 151 dyadic interactions. The annotation was performed by six independent raters recruited via Amazon Mechanical Turk, each tasked with evaluating the personality traits of individual speakers based on both the audio recordings and transcribed text of the conversations. To ensure annotation accuracy, annotators were recruited from the same sociocultural background as the original IEMOCAP speakers, i.e., native speakers of American English. The Ten Item Personality Measure (TIPI) \cite{gosling2003very}, a widely validated instrument for assessing the Big Five personality dimensions (OP, CO, EX, AG, and NE), was employed, with raters providing scores on a 7-point Likert scale (ranging from 1 = “strongly disagree" to 7 = “strongly agree") for each trait. To ensure the quality and consistency of the annotations, we implemented a rigorous data-cleaning procedure. Specifically, we excluded ratings deemed contradictory, such as instances where a speaker was simultaneously rated as “critical, quarrelsome” and “sympathetic, warm” for the AG trait, as such inconsistencies could undermine the reliability of the personality profiles. All problematic annotations were discarded, and new raters were recruited to replace them, ensuring that each conversation ultimately received six valid and independent ratings in total.  The final trait score was computed as the average of the six ratings, without applying any further normalization.

\begin{table}[tbp]
  \centering
  \caption{Inter-annotator agreement analysis for Big Five personality trait annotations. We report ICC(2,k) with 95\% confidence intervals (CI 95\%) and Fleiss' kappa to evaluate the consistency among annotators. ICC(3,k) is additionally reported as a sensitivity check.}
    \begin{tabular}{cccc}
    \toprule
    Personality traits & ICC(2,k), CI 95\% & ICC(3,k) & Fleiss' kappa \\
    \midrule
    OP & 0.89, [0.87, 0.91]  & 0.90  & 0.49 \\
    CO & 0.89, [0.87, 0.91] & 0.89  & 0.53 \\
    EX & 0.89, [0.87, 0.91] & 0.89  & 0.47 \\
    AG & 0.97, [0.96, 0.97] & 0.97  & 0.83 \\
    NE & 0.95, [0.94, 0.96] & 0.95  & 0.71 \\
    \hline
    Average & 0.92 & 0.92 & 0.61 \\
    \bottomrule
    \end{tabular}%
  \label{tab:fk}%
\end{table}%

\newcolumntype{d}[1]{D{.}{.}{#1}}
\begin{table*}[tbp]
  \centering
  \caption{Pearson correlation coefficients (PCC) between the Big Five personality traits and arousal and valence at the utterance level. Subgroup analyses are reported for male and female speakers. Values marked with $^{\dag}$ indicate a strong correlation ($|PCC|>0.5$).}
  \setlength{\tabcolsep}{8pt} 
  \renewcommand{\arraystretch}{1.2} 
  \begin{tabular}{l 
                  d{2} d{2} 
                  d{2} d{2} 
                  d{2} d{2}}
    \toprule
    & \multicolumn{2}{c}{Overall ($N=10039$)} 
    & \multicolumn{2}{c}{Male ($N=5239$)} 
    & \multicolumn{2}{c}{Female ($N=4800$)} \\
    \cmidrule(lr){2-3} \cmidrule(lr){4-5} \cmidrule(lr){6-7}
    Personality & \multicolumn{1}{c}{Valence} & \multicolumn{1}{c}{Arousal}
                & \multicolumn{1}{c}{Valence} & \multicolumn{1}{c}{Arousal}
                & \multicolumn{1}{c}{Valence} & \multicolumn{1}{c}{Arousal} \\
    \midrule
    OP & 0.53\rlap{$^{\dag}$} & 0.09 
       & 0.50\rlap{$^{\dag}$} & 0.09 
       & 0.55\rlap{$^{\dag}$} & 0.08 \\
    CO & 0.33 & -0.21 
       & 0.28 & -0.19 
       & 0.38 & -0.24 \\
    EX & 0.35 & 0.32  
       & 0.35 & 0.29   
       & 0.36 & 0.34 \\
    AG & 0.51\rlap{$^{\dag}$} & -0.11 
       & 0.46 & -0.10  
       & 0.55\rlap{$^{\dag}$} & -0.14 \\
    NE & -0.45 & 0.19  
       & -0.39 & 0.20   
       & -0.50\rlap{$^{\dag}$} & 0.19 \\
    \bottomrule
  \end{tabular}
  \label{tab:pcc_utt_subgroup}
\end{table*}

\begin{table*}[tbp]
  \centering
  \caption{Pearson correlation coefficients (PCC) between the Big Five personality traits and arousal and valence at the conversation level. Subgroup analyses are reported for male and female speakers. Values marked with $^{\dag}$ indicate a strong correlation ($|PCC|>0.5$).}
  \setlength{\tabcolsep}{8pt}
  \renewcommand{\arraystretch}{1.2}
  \begin{tabular}{l
                  d{2} d{2} 
                  d{2} d{2} 
                  d{2} d{2}}
    \toprule
    & \multicolumn{2}{c}{Overall ($N=302$)}
    & \multicolumn{2}{c}{Male ($N=151$)}
    & \multicolumn{2}{c}{Female ($N=151$)} \\
    \cmidrule(lr){2-3} \cmidrule(lr){4-5} \cmidrule(lr){6-7}
    Personality & \multicolumn{1}{c}{Valence} & \multicolumn{1}{c}{Arousal}
                & \multicolumn{1}{c}{Valence} & \multicolumn{1}{c}{Arousal}
                & \multicolumn{1}{c}{Valence} & \multicolumn{1}{c}{Arousal} \\
    \midrule
    OP & 0.70\rlap{$^{\dag}$} &  0.14
       & 0.69\rlap{$^{\dag}$} &  0.18
       & 0.71\rlap{$^{\dag}$} &  0.11 \\
    CO & 0.41  & -0.41
       & 0.33  & -0.36
       & 0.49  & -0.48 \\
    EX & 0.44  &  0.61\rlap{$^{\dag}$}
       & 0.48  &  0.58\rlap{$^{\dag}$}
       & 0.42  &  0.65\rlap{$^{\dag}$} \\
    AG & 0.63\rlap{$^{\dag}$} & -0.25
       & 0.58\rlap{$^{\dag}$} & -0.21
       & 0.68\rlap{$^{\dag}$} & -0.29 \\
    NE & -0.58\rlap{$^{\dag}$} & 0.36
       & -0.50\rlap{$^{\dag}$} & 0.36
       & -0.64\rlap{$^{\dag}$} & 0.38 \\
    \bottomrule
  \end{tabular}
  \label{tab:pcc_spk_subgroup}
\end{table*}

To evaluate the reliability of personality annotations, we employed two metrics: the intraclass correlation coefficient (ICC) \cite{fisher1970statistical} for the continuous ratings (1–7 scale) and Fleiss’ kappa \cite{fleiss2013statistical} for the binarized traits obtained via median split. Before the consistency analysis, we excluded the rating that deviated the most from the median for each trait. As shown in Table \ref{tab:fk}, inter-annotator agreement was excellent across all traits, with an average ICC(2,k) of 0.92. After data cleaning, the ICC(2,k) values ranged from 0.89 to 0.97, with 95\% confidence intervals consistently above 0.85. As a sensitivity check, we additionally report ICC(3,k), which yielded nearly identical values, confirming the robustness of the reliability estimates.
Without trimming the ratings furthest from the median score (mentioned in Section 3(A)), the average ICC(2,k) was 0.83, which still indicates strong reliability.

We also ensure annotation reliability by converting the continuous personality trait scores into binary labels using a median split, thereby maintaining robustness in simplified personality classifications (e.g., extrovert vs. introvert). The average Fleiss’ $\kappa$ across all traits was 0.61; without trimming extreme ratings, the average $\kappa$ decreased to 0.39. Among individual traits, AG and NE exhibited the highest agreement ($\kappa$=0.83 and $\kappa$=0.71, respectively), while EX had the lowest ($\kappa$=0.47), likely because the conversational context did not clearly reflect introversion–extroversion, making it harder to annotate. These results confirm the reliability and consistency of our personality annotations. The resultant annotated dataset is thereafter referred to as PA-IEMOCAP.

\subsection{Correlation of Big Five Traits and Emotions}
We introduce Pearson correlation coefficients (PCC) \cite{cohen2009pearson} between the Big Five traits and the two emotion dimensions, valence and arousal, at utterance level (Table \ref{tab:pcc_utt_subgroup}). Statistically significant correlations ($p<0.05$) were observed: OP shows strong positive correlations with valence (PCC = 0.53), which implies that individuals scoring high in OP tend to exhibit more positive emotions. NE shows a pronounced negative correlation with valence (PCC = -0.45), indicating individuals with higher NE levels express more negative emotions. AG and CO also exhibit noticeable associations with valence. Regarding arousal, EX has a positive correlation (PCC = 0.32), suggesting extroverts typically display heightened emotional energy. 

Because personality labels are constant within a conversation and thus repeated across a speaker’s utterances, treating utterances as independent may violate the independence assumption. We therefore averaged valence and arousal for each speaker in a conversation and computed PCCs across conversations (Table \ref{tab:pcc_spk_subgroup}), thereby capturing the association between personality and overall emotion expression.
Compared with the utterance-level analysis, correlations show modest increases overall; in particular, the association between Extraversion and arousal increases to (PCC=0.61). We hypothesize that aggregating at the speaker level reduces within-speaker noise, thereby yielding better correlations.

In addition, we conducted subgroup analyses by gender. The results for male and female speakers show broadly consistent patterns with the overall population, and no substantial differences between the genders were observed. For instance, the OP–valence and EX–arousal associations remained strong across both groups, while NE consistently exhibited negative correlations with valence.

\begin{figure*}[tb]
\centering
\centerline{\includegraphics[width = 0.99\linewidth]{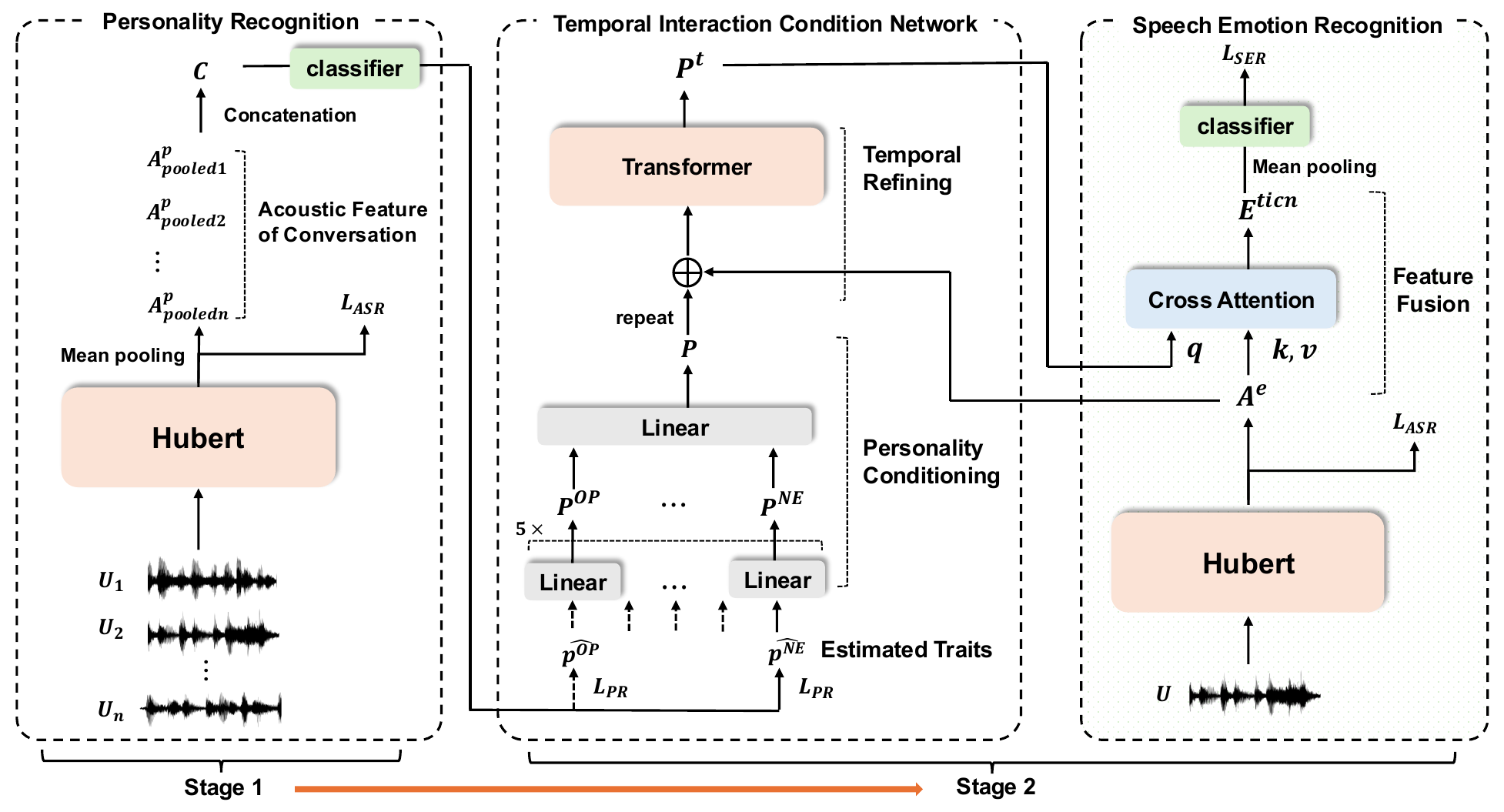}}
\vspace{10pt}
\caption{Overall flowchat of the proposed approach. We use all the utterances of a whole conversation to predict personality traits. Note that, we conduct independent experiments for prediction of each personality traits. The predicted traits are then projected by the proposed temporal interaction condition network (TICN) for improving SER.}
\label{fig:model}
\end{figure*}

\section{SER with Ground-Truth Personality Traits}
In the previous section, we confirmed a strong correlation between Big Five personality traits and emotion expression, particularly valence. 
Based on these findings, in this work, we explore the interplay between SER and PR in deep learning based models to validate the applications of these correlations in dialogue systems. 
First, we investigate whether incorporating personality information can enhance SER performance. 
Specifically, we introduce a condition network to project the Big Five personality traits into latent features. 
These personality features are then combined with acoustic features to assess whether personality information can benefit the model for emotion understanding. 

Given an utterance $U \in \{U_1, U_2, ..., U_n\}$, we incorporate HuBERT \cite{hsu2021hubert}, which consists of 7 convolutional layers followed by 12 Transformer layers as acoustic feature extractor. 
This model is pretrained in a self-supervised learning manner and can capture both acoustic features and content information from the input utterance. 
The acoustic features learned from HuBERT are represented as $A^e \in \mathbb{R}^{t \times d}$, where $t$ is the number of temporal frames and $d$ is the hidden dimension of the Transformer layers.
Previous works have validated that emotion perception requires both acoustic features and linguistic content. 
Therefore, we incorporate an ASR component to learn the linguistic information and 
thus ensure promising performance for SER \cite{cai2021speech,gao2024enhancing}. We feed $A^e$ into a linear layer and apply the connectionist temporal classification (CTC) loss function \cite{graves2006connectionist}:
\begin{equation}
\mathcal{L}_{\text{ASR}} = \text{CTC}(y, \hat{y}),
\end{equation}
where $y$ represents the ground-truth transcription, and $\hat{y}$ denotes the predicted probability sqequence.
Subsequently, we apply mean pooling across the time dimension of $A^e$ to obtain the utterance-level feature $A^e_{\text{pooled}} \in \mathbb{R}^{d}$, which is then used for either valence or arousal recognition:
\begin{equation}
    \mathcal{L}_{\text{SER}} = \left( e - \hat{e} \right)^2,
\end{equation}
where the predicted emotion \(\hat{e}\) is derived from \(A^e_{\text{pooled}}\), and \(e\) is the ground-truth emotion label. 
The overall loss function of the SER baseline integrates both ASR and SER objectives as follows:
\begin{equation}
    \mathcal{L}_{\text{SER\_baseline}} = (1- \lambda) \mathcal{L}_{\text{SER}} + \lambda \mathcal{L}_{\text{ASR}},
    \label{eq:overall_loss}
\end{equation}
where $\lambda$ is a weight parameter that balances the contribution of $\mathcal{L}_{\text{ASR}}$ and $\mathcal{L}_{\text{SER}}$. 

Inspired by the correlation between personality and emotion expression in Table \ref{tab:pcc_utt_subgroup} and \ref{tab:pcc_spk_subgroup}, we first introduce a linear layer to project the personality features $P$ from each Big Five traits $t \in \mathbb\{{OP, CO, EX, AG, NE}\}$. 
The pooled acoustic features $A^e_{\text{pooled}}$ and the personality features $P$ are then concatenated to form a combined feature $E \in \mathbb{R}^{d + e}$ for SER. We also provide the comparison of using all personality features as additional information in the experiment section.

\section{Personality Recognition and SER with predicted Personality Traits}
In real-world dialogue system applications, users do not typically provide explicit information about their personality traits. Therefore, instead of relying on ground-truth personality traits, we explore the use of predicted personality traits for improving SER.
We first conduct comprehensive PR experiments using PA-IEMOCAP, evaluating recognition performance at both the utterance and conversation levels. Given the predicted personality, our framework enables dialogue systems to infer personality traits directly from speech input.
Furthermore, we introduce a TICN to capture the temporal dynamics of personality impact on emotional expression and incorporate cross-attention for effective feature fusion (Fig. \ref{fig:model}).

\subsection{Utterance-level Personality Recognition}
To align with SER, we first introduce utterance-level PR.
First, we implement utterance-level PR by employing the HuBERT for acoustic feature extraction. 
Given the significant role of linguistic content in this task, we integrate an ASR component similar to our SER framework, and we extract the utterance-level feature $A^p_{pooled}$ through mean pooling for PR with the following loss function:
\begin{equation}
    \mathcal{L}_{\text{PR}} =  \left( p - \hat{p} \right)^2,
\end{equation}
where $p \in \mathbb\{{p^{OP}, p^{CO}, p^{EX}, p^{AG}, p^{NE}}\}$ is the ground-truth personality traits, and $\hat{p}$ represents the corresponding predicted personality traits. 
The overall loss function for our PR baseline combines both PR and ASR objectives:
\begin{equation}
    \mathcal{L}_{\text{PR\_baseline}} = (1- \gamma) \mathcal{L}_{\text{PR}} + \gamma \mathcal{L}_{\text{ASR}},
\end{equation}
where $\gamma$ is a weight parameter balancing the contribution of these two tasks.

Second, building upon our separate SER and PR frameworks, we explore multi-task learning to jointly train both tasks at the utterance-level to explore the mutual influence of these two tasks. Given an input utterance $U$, we extract acoustic features $A^p$ for three objectives: ASR, SER, and PR.
For the ASR component, we maintain the same approach and apply the CTC loss function.
After applying mean pooling across the time dimension to obtain $A^p_{\text{pooled}} \in \mathbb{R}^{d}$, we feed this utterance-level representation for both SER and PR.
The overall loss function for our multi-task learning framework integrates all three objectives:
\begin{equation}
\mathcal{L}_{\text{multitask}} =  (1-\alpha-\beta) \mathcal{L}_{\text{PR}} + \alpha \mathcal{L}_{\text{SER}} + \beta \mathcal{L}_{\text{ASR}},
\end{equation}
where $\alpha$ and $\beta$ are weight parameters that balance the contribution of each task, subject to the constraint $0 \leq \alpha, \beta \leq 1$ and $\alpha + \beta \leq 1$.

\subsection{Conversational-level Personality Recognition
}
We extend to conversation-level PR, acknowledging that personality traits become more apparent across longer conversations. 
For a conversation consisting of multiple utterances unit $\{U_1, U_2, ..., U_n\}$, we extract utterance-level acoustic features and then concatenate them as conversational-level feature:
\begin{equation}
    C^p = Concat[A_{pooled1}^{p}; A_{pooled2}^{p}; ...; A_{pooledn}^{p}],
\end{equation}
where $A_{pooledj}^{p}$ denotes the utterance-level acoustic features for the $j$-th utterance. The conversation-level representation $C$ captures the holistic personality manifestation across multiple utterances, enabling more robust PR in dialogue systems.
The automatic PR component serves as a basis for our subsequent exploration of how predicted personality traits can enhance SER.

\subsection{Temporal Interaction Condition Network (TICN)}
Self-attention is shown to be effective for SER as emotional expression varies at different temporal segments of an utterance. Although the personality of a speaker remains consistent, we hypothesize that its influence on emotion expression varies across different temporal segments of an utterance. To enable the model to capture this dynamic alignment, we replicate $P$ across time steps, not to imply temporal variability in personality itself, but to allow personality information to interact with acoustic features at each segment. Through TICN, this design facilitates the extraction of fine-grained personality-conditioned features that reflect the varying salience of personality traits in shaping emotional expression over time. Specifically, we unsqueeze the personality feature $P$ projected from linear layer as $P^t \in \mathbb{R}^{t \times d}$ to align their dimensions with $A^e$ (acoustic feature for SER). We integrate the temporal emotion information of the input utterance by performing element-wise addition between $A^e$ and $P$, which can be denoted as:
\begin{equation}
P = 0.9 \cdot P \oplus 0.1 \cdot A^e 
\end{equation}
Finally, these fused features are passed through a Transformer layer to learn the interaction of personality and emotion expression over time.
To effectively integrate the finegrained personality features with acoustic features, we apply a cross attention mechanism, allowing the model to adaptively capture relevant interactions between the two input features: 
\begin{equation}
\setlength{\abovedisplayskip}{4pt}
	\setlength{\belowdisplayskip}{4pt}
       E^\mathrm{ticn} = \text{softmax}\left( (P^t W^Q) (A^e W^K)^T / \sqrt{d_k} \right) (A^e W^V) 
\end{equation}
Here, $P^t$ serve as the query, while $A^e$ act as both the key and the value, with the number of attention heads set to 4. We also examined the reverse configuration, in which $P^t$ acted as key/value and $A^e$ as query; however, this design yielded unsatisfactory results and was not further considered. This proposed approach enables $A^e$ to selectively incorporate information from $P^t$ that is most relevant to emotional expressions at each temporal segment. 

As shown in the right sub-figure of Fig. \ref{fig:model}, the outputs of the cross-attention layer $E^\mathrm{ticn} \in \mathbb{R}^{t \times d}$ are then pooled to obtain fixed-length representations, which are subsequently used for SER.
The overall loss function of SER follows the same formulation as in Eq. (\ref{eq:overall_loss}). 

\begin{table*}[!t]
\renewcommand{\arraystretch}{1.2}
\centering
\caption{Effect of ground-truth Big Five traits in arousal and valence recognition. We use ``*'' to denote statistically significant improvement (p $<$ 0.05) from the baseline.}
\begin{tabular}{ccccc|ccc|ccc}
\toprule
\multicolumn{5}{c|}{Model}                                                                                          & \multicolumn{3}{c|}{Valence (CCC)}                                        & \multicolumn{3}{c}{Arousal (CCC)}                                        \\
\midrule
\multicolumn{5}{c|}{Baseline}                                                                                       & \multicolumn{3}{c|}{0.698}                                          & \multicolumn{3}{c}{0.711}                                          \\
\midrule
\multicolumn{1}{c}{OP} & CO                   & EX                   & AG                   & NE                   & Concat.              & CA                   & TICN-CA               & Concat.              & CA                   & TICN-CA               \\
\midrule
\ding{51} & & & & & \, 0.737* & \, 0.755* & \, 0.782* & 0.707 & 0.707 & 0.707 \\
& \ding{51} & & & & 0.716 & \, 0.739* & \, 0.751* & 0.713 & 0.708 & 0.718 \\
& & \ding{51} & & & 0.712 & \, 0.745* & \, 0.740* & 0.719 & \textbf{0.720} & 0.716 \\
& & & \ding{51} & & \, 0.733* & \, 0.754* & \, 0.760* & 0.709 & 0.711 & 0.708 \\
& & & & \ding{51} & 0.726 & \, 0.750* & \, 0.768* & 0.712 & 0.720 & 0.715 \\
\midrule
\ding{51} & \ding{51} & \ding{51} & \ding{51} & 
\ding{51} & \, 0.746* & \, 0.770* & \, \textbf{0.785}* & 0.707 & 0.717 & 0.711 \\
\bottomrule
\end{tabular}
\label{tab:groundtruth}
\end{table*}
\begin{figure*}[!t]
    \centering
    \begin{minipage}[b]{0.32\textwidth}
        \centering
        \includegraphics[width=5.4cm, height=4.3cm]{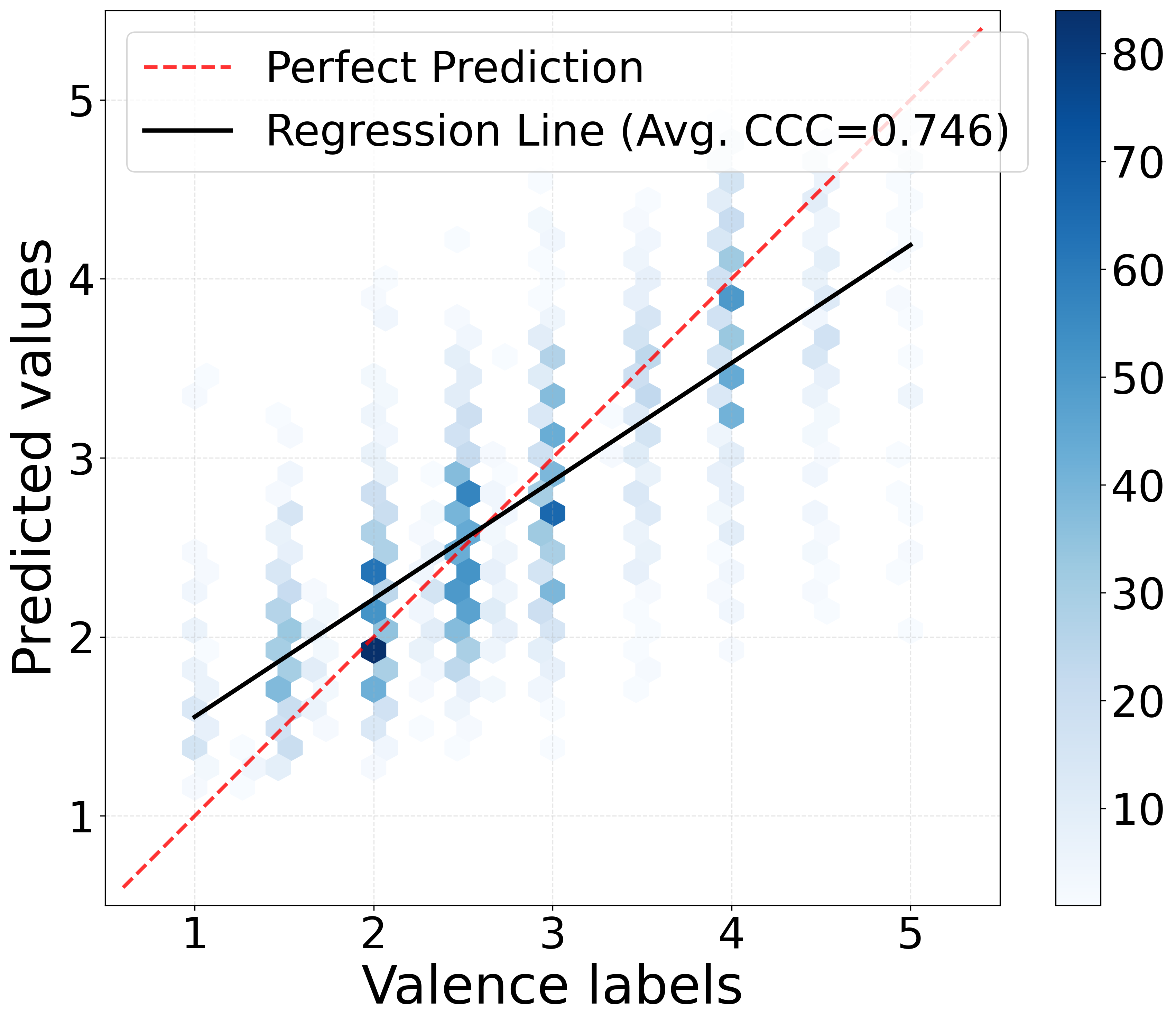}
        \caption*{(a) Concat.}
        \label{fig:a}
    \end{minipage}
    \hfill
    \begin{minipage}[b]{0.32\textwidth}
        \centering
        \includegraphics[width=5.4cm, height=4.3cm]{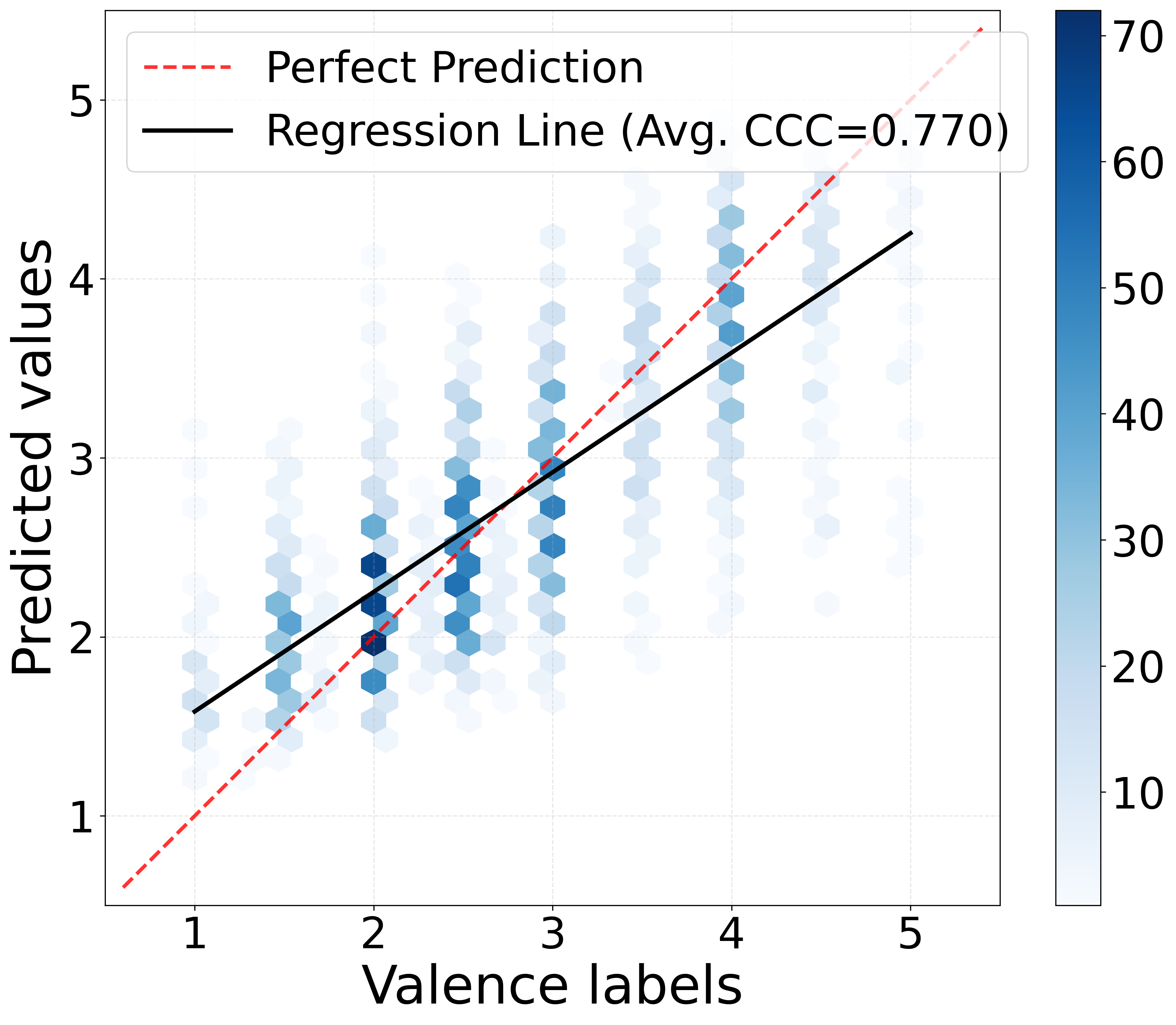}
        \caption*{(b) CA}
    \end{minipage}
    \hfill
    \begin{minipage}[b]{0.32\textwidth}
        \centering
        \includegraphics[width=5.4cm, height=4.3cm]{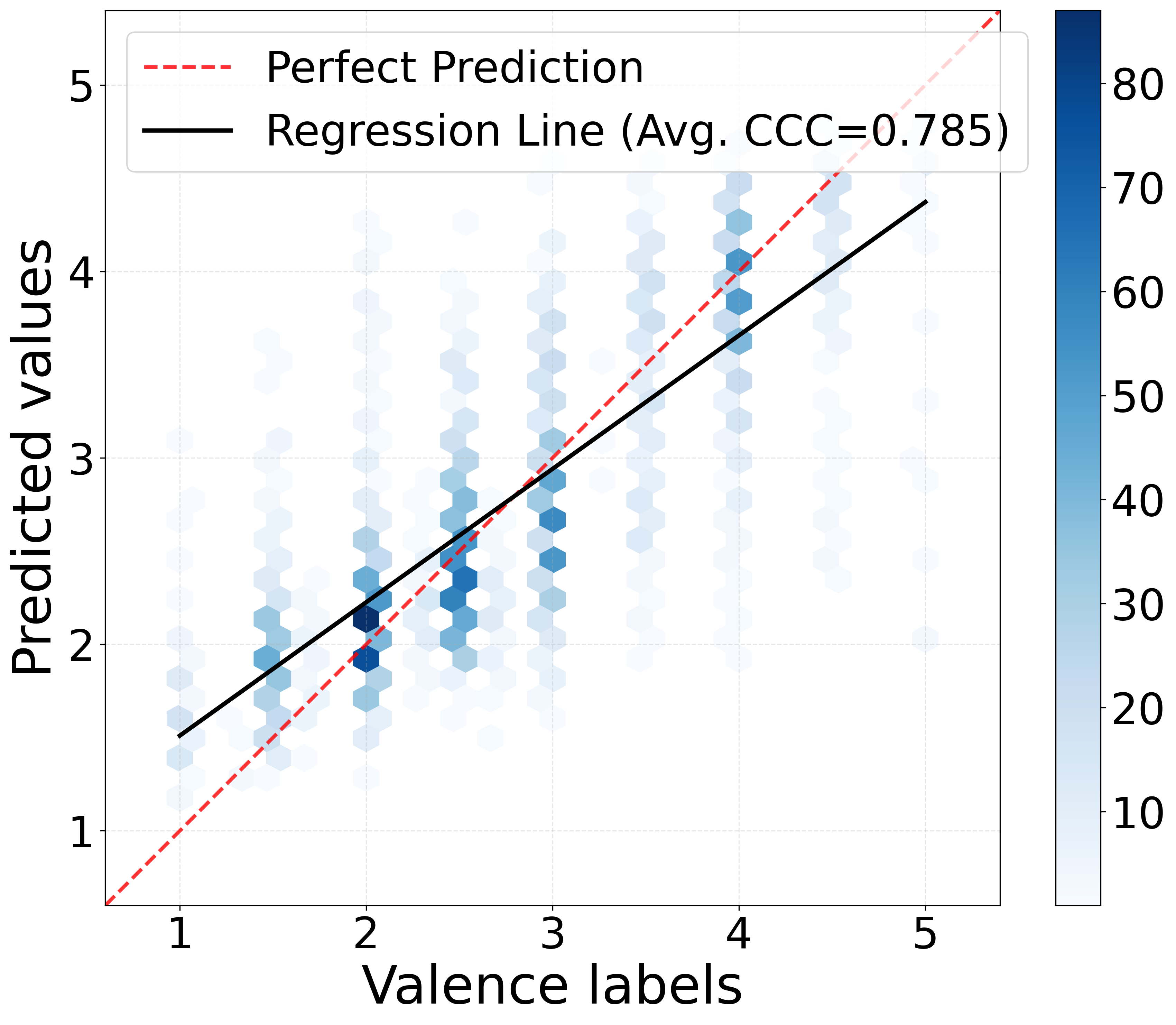}
        \caption*{(c) TICN-CA}
    \end{minipage}
    \caption{Hexbin density plot of valence recognition using different condition network integrating all ground-truth Big Five traits.}
    \label{fig:gtfig}
\end{figure*}

\section{Experimental Evaluation}
This section systematically examines the influence of personality information on SER. To comprehensively assess this relationship, we conducted three distinct experiments. The first experiment evaluates the effect of ground-truth personality traits in enhancing SER performance. This is achieved by integrating personality traits into SER systems through three distinct approaches. 
The second component investigates PR across multiple contexts. We first establish a baseline by performing utterance-level PR. Then, we implement multi-task learning to simultaneously optimize for SER and PR, examining the mutual effects between these tasks. Considering that personality traits typically remain consistent throughout a conversation, we expand our analysis to conversational-level PR, allowing us to better capture the temporal dynamics inherent in entire conversations.
The third component explores practical applications by utilizing predicted personality traits to improve SER performance. This experiment adopts the same condition network for Big Five traits as those employed in the first experiment, ensuring consistency across our evaluations.
\subsection{Implementation and Setup}
We implemented the proposed model using PyTorch and the Huggingface Transformers repository \cite{wolf2020Transformers}. 
Our experiments utilize HuBERT-base \cite{hsu2021hubert} as the acoustic feature extractor, which was pretrained on 60,000 hours of Libri-Light data \cite{kahn2020libri} and comprises a convolutional feature extractor followed by 12 Transformer encoder layers with 768-dimensional hidden representations. For the linear layers in TICN, the output dimension was fixed at 768 to ensure consistency between $P^t$ and $A^e$. During training for both SER and PR tasks, the CNN feature extractor together with the first six Transformer layers were frozen, while the remaining six layers were fine-tuned. Optimization was performed using AdamW with a learning rate of $5 \times 10^{-5}$, a mini-batch size of 2, and gradient accumulation over 8 mini-batches. The model was trained with dynamic batch padding to accommodate varying input lengths. For multi-task learning experiments, the weight parameters $\lambda$, $\gamma$, $\alpha$, and $\beta$ were tuned among ${1, 0.1, 0.01}$, with 0.1 yielding the best performance for all of them and thus adopted in our experiments. Each experiment was repeated three times with the default random seed of Huggingface and we report the average results.

We split the PA-IEMOCAP dataset into a training set (sessions 2–4), a validation set (14 conversations from session 1), and a test set (another 14 conversations from session 1), ensuring speaker-independent evaluation for both SER and PR. 
We employed concordance correlation coefficient (CCC) \cite{lawrence1989concordance} as the primary metric for both PR and SER, with model checkpoints saved based on the best CCC performance (SER) on the validation set. 
\subsection{Effect of Ground-Truth Personality Traits for SER}
In this study, we explored three approaches to integrate personality information with acoustic features for enhancing SER. The first method, Concat, involves conditioning personality features through a linear layer and concatenating them with acoustic features. The second method, CA, employs a cross-attention mechanism to combine personality and acoustic features, facilitating effective interactions between the two modalities. The third method, TICN-CA, incorporates TICN to model the temporal dynamics of personality traits, which are then integrated with acoustic features via cross-attention (CA). We compare the proposed approaches with baseline model that uses only speech input. The results are presented in Table \ref{tab:groundtruth}.

We observe that concatenating acoustic features with personality embeddings significantly improved valence recognition, particularly when incorporating features learned from OP and AG. This aligns with the correlation analysis, where OP and AG exhibited the strongest correlations with valence. CA facilitates more effective interactions between personality and acoustic features, leading to noticeable improvements. The proposed TICN captures the temporal dynamics of personality traits' effects on emotional expression, thereby significantly improving valence recognition. According to Table \ref{tab:groundtruth}, the best performance was achieved by using TICN to project all five traits collectively, which were then combined with acoustic features via cross-attention. This approach yielded a CCC of 0.791 for valence recognition. The proposed model introduces only a few additional linear layers, one Transformer block, and a cross-attention mechanism, resulting in a relatively small parameter increase of about 12.1\% (106.8M vs. 95.4M in the baseline). These findings suggest that integrating all personality traits provides a more robust improvement in valence recognition compared to utilizing any single trait individually. However, improvements in arousal recognition remained limited across all personality traits, indicating that personality information show more influential impact in valence rather than arousal in emotional expressions.

Fig. \ref{fig:gtfig} shows Hexbin density plots for valence recognition, where darker hexagons indicate higher prediction density. Since sample sizes differ across valence labels, density comparisons should be made vertically along the true-label axis. For example, in plot (a), although samples at the lowest valence appear lighter overall than middle valence samples, they have darker hexagons closer to the diagonal (perfect prediction, shown in red). In contrast, samples with middle valence labels (around 3) show a relatively even vertical distribution, indicating poorer prediction accuracy. The TICN-CA approach (c) clearly provides better predictions than the concatenation model (a). Its regression line aligns more closely with the diagonal, and its data points are more densely clustered near the ideal prediction line compared to both (a) and (b). Notably, TICN-CA shows improved prediction clusters for valence values between 2 and 3, a range typically challenging for valence recognition systems, representing a significant advancement over the baseline model (a).

\begin{table}[t]
  \centering
  \caption{Experimental results of personality recognition for each trait. Results are reported using concordance correlation coefficient (CCC). Utt: Utterance-level experiment. Conv: Conversational-level experiment.}
    \begin{tabular}{cccc}
    \toprule
    Experimental settings & Setting 1 & Setting 2 & Setting 3 \\
    \midrule
    Training & Utt   & Utt   & Conv \\
    Inference & Utt   & Conv  & Conv \\
    \midrule
    OP    & 0.441 & 0.492 & \textbf{0.695} \\
    CO    & 0.333 & 0.368 & \textbf{0.748} \\
    EX    & 0.437 & 0.467 & \textbf{0.809} \\
    AG    & 0.547 & 0.597 & \textbf{0.843} \\
    NE    & 0.438 & 0.464 & \textbf{0.793} \\
    \midrule
    Avg.  & 0.439 & 0.478 & \textbf{0.778} \\
    \bottomrule
    \end{tabular}%
  \label{tab:pr}%
\end{table}%

\begin{figure*}[!t]
    \centering
    \begin{minipage}[b]{0.425\textwidth}
        \centering
        \includegraphics[width=\textwidth]{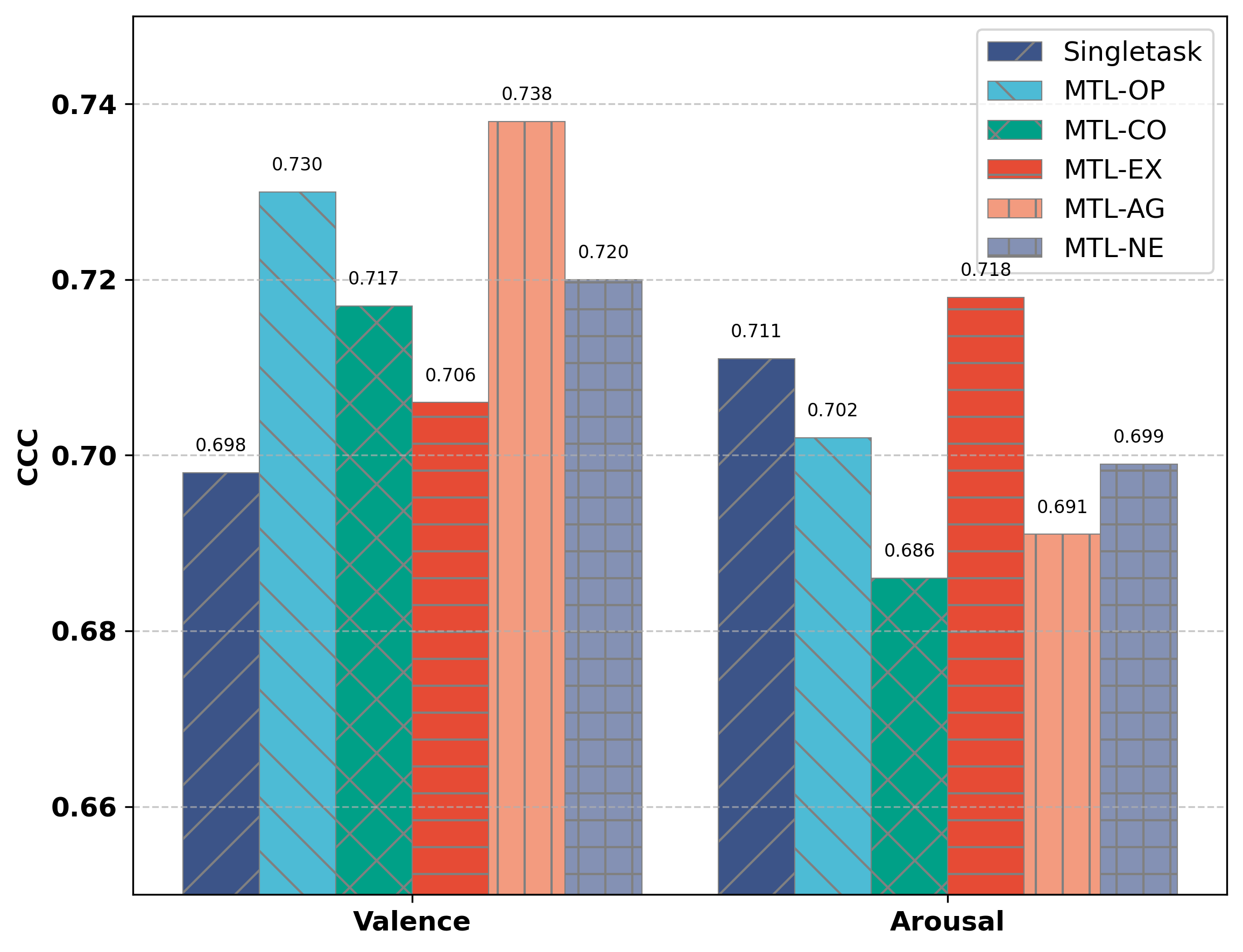}
        \caption*{(a) Emotion recognition results}
        \label{fig:emotion}
    \end{minipage}
    \hfill
    \begin{minipage}[b]{0.55\textwidth}
        \centering
        \includegraphics[width=\textwidth]{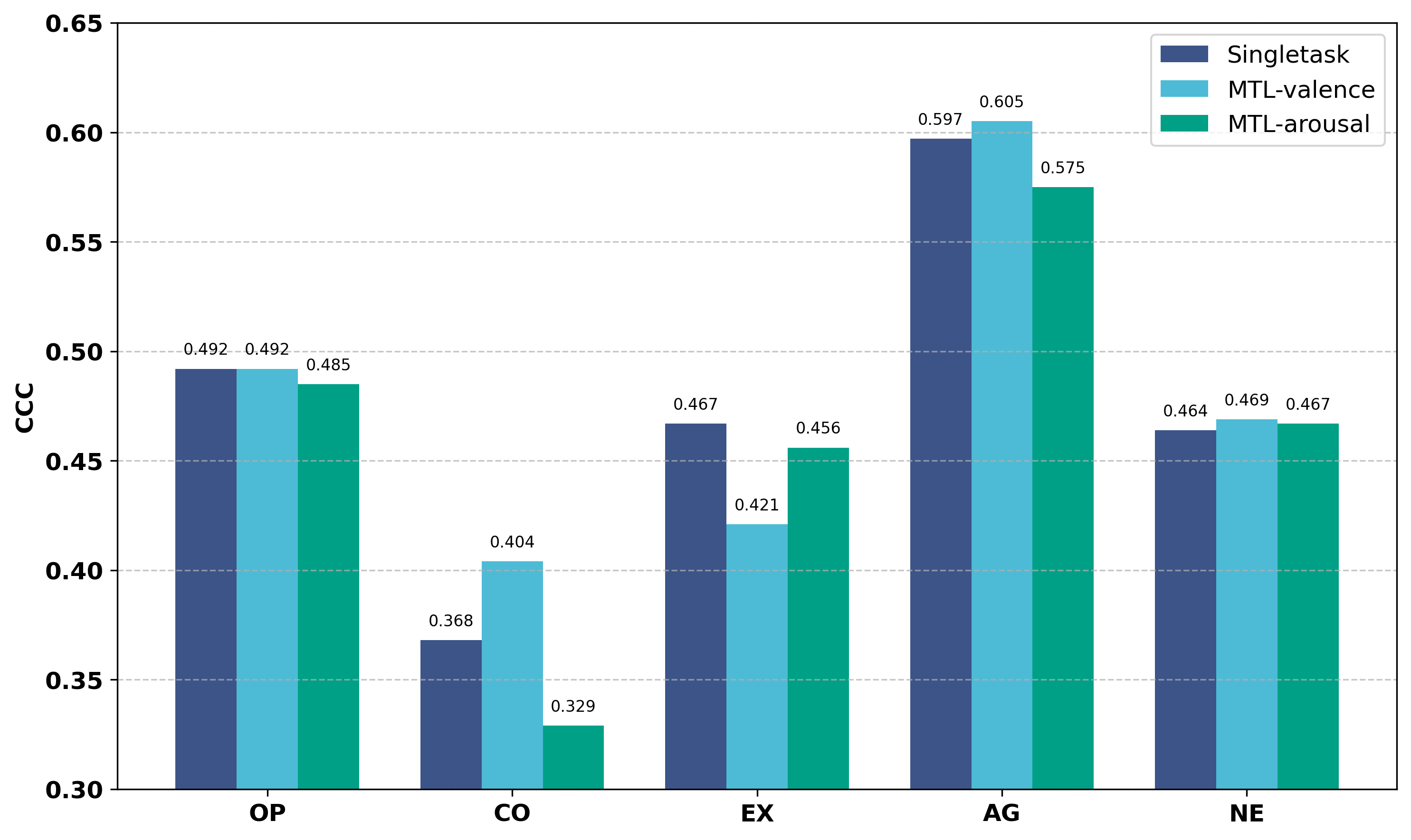}
        \caption*{(b)  Personality recognition results}
        \label{fig:personality}
    \end{minipage}
    \caption{Comparison between single-task and multi-task learning on personality and emotion recognition. Results are reported using concordance correlation coefficient (CCC).}
    \label{fig:mtl}
\end{figure*}
\subsection{Personality Recognition}
\subsubsection{Utterance-level Personality Recognition}
We develop a baseline system for PR using the PA-IEMOCAP dataset. Two experimental settings were considered. In setting 1, utterance-level predictions were performed by extracting acoustic features from each utterance using HuBERT, followed by direct estimation of personality traits. In setting 2, a postprocessing step was introduced since personality traits remain consistent throughout a conversation. Specifically, predictions from all utterances belonging to the same speaker within a conversation were averaged to produce a single prediction per speaker. The CCC was then computed using these aggregated predictions.

The results of the PR experiments are presented in Table \ref{tab:pr}. We observe that AG gets the best recognition performance, and CO is the most difficult to predict. This indicates that individual utterances contain personality information, and achieved moderate CCC across all Big Five traits. 
Applying post-processing by averaging predictions across all utterances from the same speaker, we observed consistent improvements for all five traits. The CCC values increased by an average of 0.048, with the most significant improvement observed for AG and OP. 
This result validates that postprocessing mitigates the variability inherent in utterance-level experiments, and personality traits manifest more consistently across multiple utterances.
\subsubsection{Conversational-Level Personality Recognition}
Personality traits remain consistent throughout an entire conversation, although their expression may vary across individual utterances. Consequently, a conversational-level approach can yield more accurate predictions of the Big Five traits. By leveraging all utterances, this method allows the model to estimate the personality traits considering the whole conversation, and achieve the best performance. In this section, we conducted conversational-level experiments by concatenating the features extracted from HuBERT for all utterances of each speaker (setting 3).

As shown in the rightmost column of Table \ref{tab:pr}, this approach significantly outperformed utterance-level analysis, with CCC improvements exceeding 0.2 across all traits. AG achieves the best performance, with a CCC of 0.843, while OP exhibites the smallest improvement, achieving 0.695 on CCC. 

These results validate that predicting personality traits should take into account information from the entire conversation, as conversation-level features effectively capture contextual variations. The significant improvement between these two settings underscores the importance of broader conversational context and speaking patterns, suggesting that, much like human perception, computational models also require sufficient conversational evidence to form a reliable impression of a speaker’s personality.

\subsubsection{Multi-task Learning for Speech Emotion Recognition and Personality Recognition}
As a common practice of leveraging emotional-related information, we employ a simple multi-task learning for SER and PR to investigate the mutual influence between these two tasks. Specifically, we incorporate HuBERT as feature encoder, and the output feature are used for both SER and PR. Since SER is inherently conducted at the utterance-level, we incorporate utterance-level acoustic features for both SER and PR. Previous experiments have validated that personality information can benefit SER, motivating us to explore how these tasks can benefit through joint training.

The experimental results are shown in Fig. \ref{fig:mtl}. For valence recognition, multi-task learning consistently outperforms the baseline across all personality traits, with improvements ranging from 0.008 to 0.04. In contrast, arousal recognition performance was degraded from the baseline except for EX.
For PR, no improvement is observed, except for CO with valence recognition. These results suggest that simple multi-task learning cannot effectively improve SER and PR.

\subsection{Conditioning Predicted Personality Traits for SER}
\begin{table*}[t]
\renewcommand{\arraystretch}{1.2}
  \centering
  \caption{Valence recognition performance when conditioning on predicted or ground-truth personality traits using TICN-CA. We evaluate how using different Big Five traits (left) and different PR approaches (middle) affects valence recognition. Results are reported using concordance correlation coefficient (CCC). We use ``*'' to denote statistically significant improvement (p $<$ 0.05) from the baseline.}
    \begin{tabular}{ccccc|ccc|c}
    \toprule
    \multicolumn{5}{c|}{Incorporated Traits} & \multicolumn{3}{c|}{Predicted Traits (PR approach)} & ground-truth \\
    \midrule
    OP    & CO    & EX    & AG    & NE    & Setting 1 & Setting 2 & Setting 3 & Oracle \\
    \midrule
    \ding{51} &       &       &       &       & 0.701 & \, 0.749* & \, 0.756* & \, 0.782* \\
          & \ding{51} &       &       &       & 0.688 & 0.707 & 0.728 & \, 0.751* \\
          &       & \ding{51} &       &       & 0.692 & 0.693 & 0.735 & \, 0.740* \\
          &       &       & \ding{51} &       & 0.710 & 0.729 & \, 0.761* & \, 0.760* \\
          &       &       &       & \ding{51} & 0.705 & 0.723 & \, 0.762* & \, 0.768* \\
    \midrule
    \ding{51} & \ding{51} & \ding{51} & \ding{51} & \ding{51} & 0.695 & \, 0.754* & \, \textbf{0.776*} & \, \textbf{0.785*} \\
    \bottomrule
    \end{tabular}
  \label{tab:predgt}
\end{table*}
\begin{figure*}[t]
    \centering
    \begin{minipage}[b]{0.3\textwidth}
        \centering
        \includegraphics[width=6cm, height=5cm]{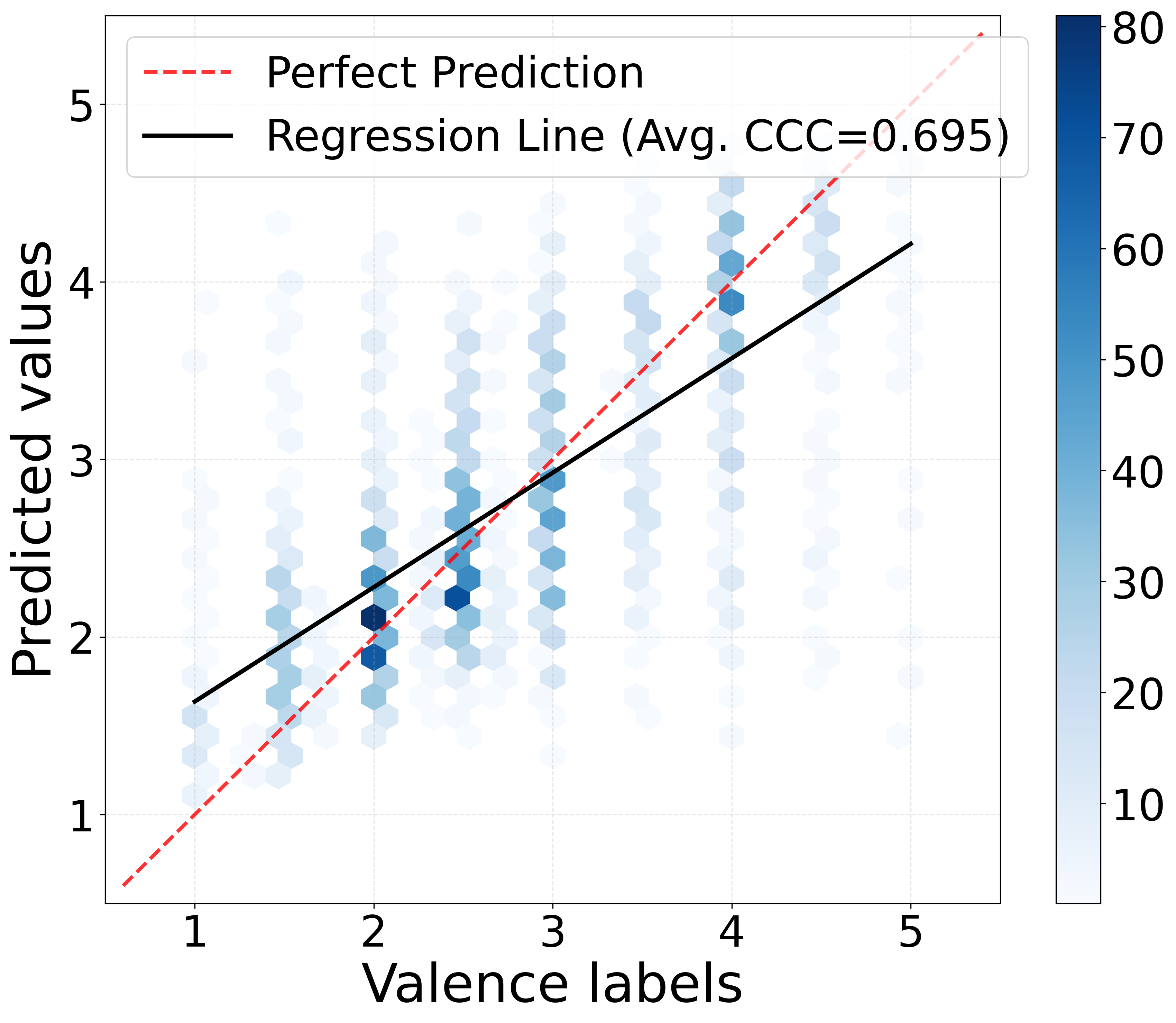}
        \caption*{(a) Setting 1}
        \label{fig:a}
    \end{minipage}
    \hfill
    \begin{minipage}[b]{0.3\textwidth}
        \centering
        \includegraphics[width=6cm, height=5cm]{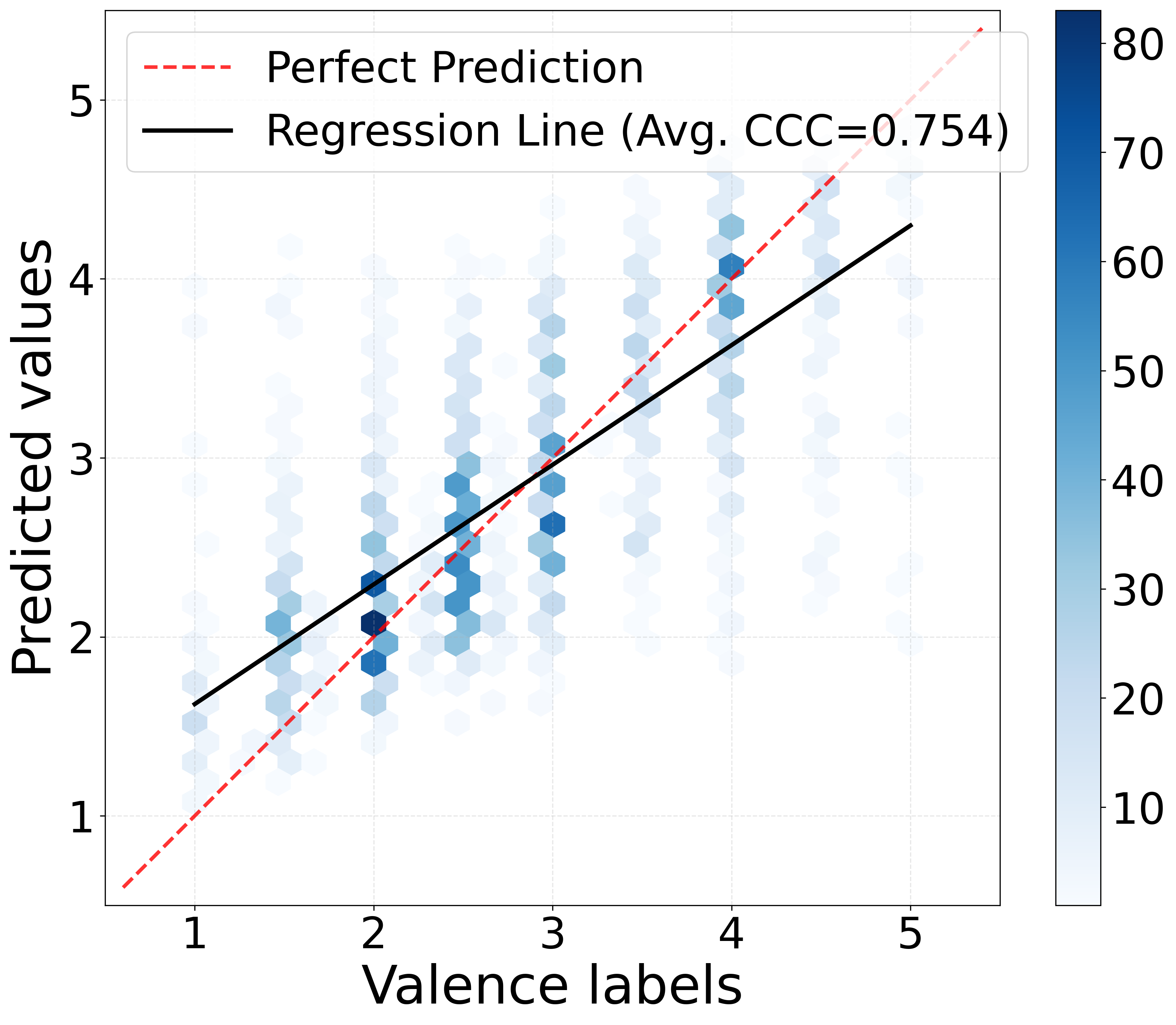}
        \caption*{(b) Setting 2}
    \end{minipage}
    \hfill
    \begin{minipage}[b]{0.3\textwidth}
        \centering
        \includegraphics[width=6cm, height=5cm]{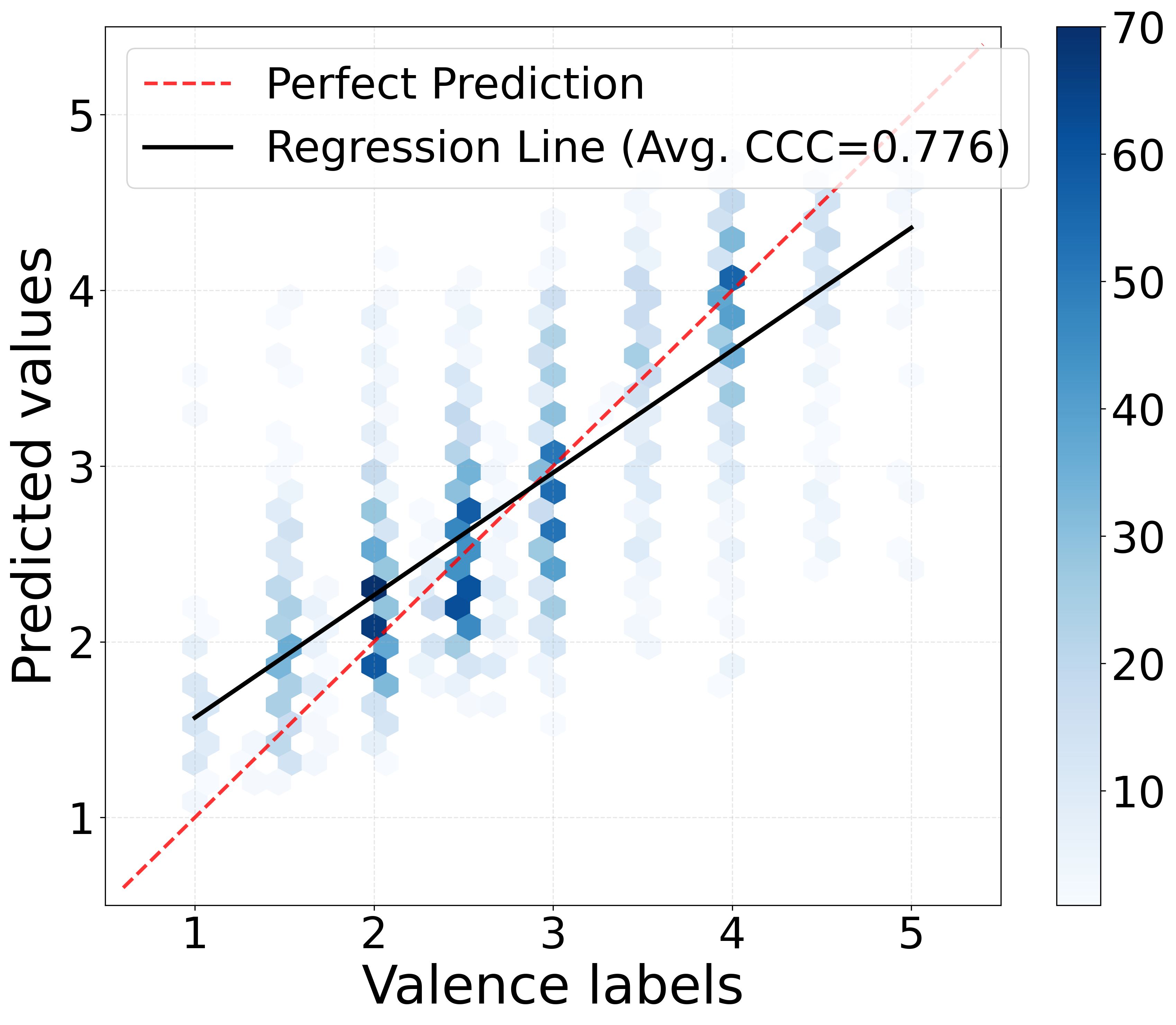}
        \caption*{(c) Setting 3}
    \end{minipage}
    \caption{Hexbin density plot of valence recognition using different condition networks integrating predicted personality traits. We compare three different settings estimating the personality traits.}
    \label{fig:predfig}
\end{figure*}
The results of Table \ref{tab:groundtruth} and Fig. \ref{fig:mtl} (a) have shown that incorporating personality information can benefit valence recognition. In this section, we conduct experiments using predicted personality labels and more sophisticated TICN for enhancing SER. The predicted labels were obtained from the three different personality recognition settings described in the previous section. To integrate the personality information, we employed TICN-CA, which showed the best performance in Table \ref{tab:groundtruth}. Table \ref{tab:predgt} summarizes the results.

These results indicate that incorporating predicted personality traits improves valence recognition, with the conversational-level PR approach yielding the highest performance across all settings. When individual traits were incorporated, OP, AG, and NE led to the most notable improvements, consistent with earlier findings. The integration of all five traits predicted by conversational-level PR achieved the best results. In contrast, incorporating traits predicted from single utterance showed limited performance. These results further validate the effectiveness of personality-aware SER and highlight the advantages of leveraging conversational context for more reliable personality trait estimation. Notably, the introduction of the conversational PR module increases the parameter count substantially (from 95.4M in the baseline SER model to 333.9M in total), which should be considered as the trade-off for the performance gain.

Fig. \ref{fig:predfig} shows hexbin density plots of valence recognition results using the predicted personality traits from the three settings. Compared with setting 1, we observe notably improved distribution in both setting 2 and 3, even though using postprocessing at the inference stage leads to modest improvement in PR performance. 

\begin{table}[tbp]
  \centering
  \caption{Different levels of Gaussian noise injected into the personality labels during inference. Results are reported using concordance correlation coefficient (CCC).}
    \begin{tabular}{cc}
    \toprule
    Gaussian noise & Valence (CCC) \\
    \midrule
    0.0 (GT)     & 0.785 \\
    \midrule
    0.1   & 0.782 \\
    0.2   & 0.780 \\
    0.4   & 0.778 \\
    0.8   & 0.743 \\
    1.2   & 0.697 \\
    1.6   & 0.657 \\
    \bottomrule
    \end{tabular}%
  \label{tab:noise}%
\end{table}%

In practical applications, a valence recognition model may be trained with ground-truth personality information but must rely on automatically predicted traits during inference, whose accuracy gradually improves as the dialogue progresses. To examine robustness under such conditions, we injected zero-mean Gaussian noise with varying standard deviations into the personality labels at inference and evaluated its impact on valence recognition. As shown in Table~\ref{tab:noise}, the model remains stable when the noise standard deviation is below 0.8, achieving results comparable to those obtained with ground-truth personality labels (0.785 vs. 0.743). This finding indicates that even coarse or approximate personality estimates are sufficient to provide meaningful benefits for valence recognition. Performance declines more noticeably under larger perturbations (e.g., CCC = 0.657 at $\sigma=1.6$), yet the results demonstrate that the model maintains considerable robustness to moderate levels of personality recognition error.
\subsection{Comparasion with existing approaches}
To facilitate comparison with prior studies, we also report the SER results under the leave-one-session-out (LOSO) data split. 

\begin{table}[tbp]
  \centering
  \caption{Comparison with existing valence recognition approaches.}
    \begin{tabular}{lccc}
    \toprule
    Approach & Input & Year  & CCC \\
    \midrule
    Srinivasan \cite{srinivasan2022representation} & Speech & 2022  & 0.582 \\
    Wagner \cite{wagner2023dawn} & Speech & 2023  & 0.478 \\
    Ispas \cite{ispas2023multi} & Speech \& Text & 2023  & 0.744 \\
    Vlasenko \cite{vlasenko2024comparing} & Speech & 2024  & 0.683 \\
    Zhou \cite{zhou2024learning} & Speech & 2024  & 0.674 \\
    Messaoudi \cite{messaoudi2024modeling}& Text  & 2024  & 0.724 \\
    \midrule
    Baseline (w/o traits) & Speech & 2025  & 0.681 \\
    Proposed (Predicted traits) & Speech & 2025  & 0.736 \\
    Proposed (GT traits) & Speech & 2025  & \textbf{0.766} \\
    \bottomrule
    \end{tabular}%
  \label{tab:sota}%
\end{table}%
The results in Table~\ref{tab:sota} demonstrate the rapid progress of valence recognition systems over recent years. Notably, text-based models (e.g., \cite{messaoudi2024modeling}) generally outperform speech-based counterparts, suggesting that lexical cues provide highly informative signals for valence estimation. Furthermore, multimodal approaches that jointly leverage speech and text (e.g., \cite{ispas2023multi}) have achieved the highest reported CCC scores in prior studies, highlighting the complementary nature of linguistic and acoustic information in capturing valence.

As shown in Table \ref{tab:sota}, our baseline system achieved performance comparable to other speech-based models by incorporating ASR as an auxiliary task. Incorporating ground-truth (GT) personality traits yielded the best performance, with a CCC of 0.766. When conditioning on predicted personality labels, our model also surpassed both existing speech- and text-based approaches and achieved comparable performance with multimodal system using both speech and text. It is important to note that text input was used only during training for ASR; during inference, the model relied solely on speech. Therefore, the most appropriate comparison is with other speech-based systems.

\section{Conclusions}
This study investigates the intrinsic relationship between personality traits and emotional expression in speech, especially exploring how personality information can be effectively leveraged to improve speech emotion recognition (SER). We annotated the well-known IEMOCAP emotional dataset with personality labels and identified strong correlations between traits and emotional expression, particularly valence. Then we present a novel approach for personality-aware SER.

We first established comprehensive baselines for PR at both utterance and conversation levels. The conversation-level models consistently outperformed utterance-level models by over 0.2 concordance correlation coefficient (CCC) across all five personality traits, reflecting the stability of personality characteristics over extended interactions.
To capitalize on the strong correlation between personality traits and emotional expression, we propose the temporal interaction condition network with cross attention for feature fusion (TICN-CA). The proposed approach is designed to capture the dynamic impact of predicted personality traits on emotional expression across temporal segments of speech. Our experimental results demonstrate that integrating personality traits, whether ground-truth or predicted, significantly improves SER performance, particularly for valence. Using ground-truth personality traits, TICN improved the CCC from 0.698 to 0.785, while using predicted traits from conversation-level PR experiments as input achieved a comparable CCC of 0.776.

These findings validate the efficacy of personality-aware SER systems and highlight their potential for real-world applications where explicit personality data is unavailable. By demonstrating that predicted personality traits can substantively enhance emotion recognition, our approach paves the way for more personalized and emotionally intelligent human-computer interaction systems. This work not only addresses the gap in available emotion and personality annotated datasets but also establishes a foundation for future research into multimodal and context-aware affective computing. While the personality annotations were derived from acted performances, we also aim to extend our study to more naturalistic data in future work.
\section{Acknowledgment}
This work was supported by JST SPRING (JPMJSP2110), and JST Moonshot R$\&$D (JPMJPS2011).

\bibliographystyle{IEEEtran}
\bibliography{refs}

@inproceedings{li2019improved,
  title={Improved End-to-End Speech Emotion Recognition Using Self Attention Mechanism and Multitask Learning.},
  author={Li, Yuanchao and Zhao, Tianyu and Kawahara, Tatsuya and others},
  booktitle={Interspeech},
  pages={2803--2807},
  year={2019}
}

@inproceedings{gat2022speaker,
  title={Speaker normalization for self-supervised speech emotion recognition},
  author={Gat, Itai and Aronowitz, Hagai and Zhu, Weizhong and Morais, Edmilson and Hoory, Ron},
  booktitle={ICASSP 2022-2022 IEEE International Conference on Acoustics, Speech and Signal Processing (ICASSP)},
  pages={7342--7346},
  year={2022},
  organization={IEEE}
}

@inproceedings{pappagari2020x,
  title={x-vectors meet emotions: A study on dependencies between emotion and speaker recognition},
  author={Pappagari, Raghavendra and Wang, Tianzi and Villalba, Jesus and Chen, Nanxin and Dehak, Najim},
  booktitle={ICASSP IEEE International Conference on Acoustics, Speech and Signal Processing (ICASSP)},
  pages={7169--7173},
  year={2020},
  organization={IEEE}
}

@article{cowie2001emotion,
  title={Emotion recognition in human-computer interaction},
  author={Cowie, Roddy and Douglas-Cowie, Ellen and Tsapatsoulis, Nicolas and Votsis, George and Kollias, Stefanos and Fellenz, Winfried and Taylor, John G},
  journal={IEEE Signal processing magazine},
  volume={18},
  number={1},
  pages={32--80},
  year={2001},
  publisher={IEEE}
}

@inproceedings{sharma2022multi,
  title={Multi-lingual multi-task speech emotion recognition using wav2vec 2.0},
  author={Sharma, Mayank},
  booktitle={ICASSP 2022-2022 IEEE International Conference on Acoustics, Speech and Signal Processing (ICASSP)},
  pages={6907--6911},
  year={2022},
  organization={IEEE}
}

@inproceedings{wolf2020transformers,
  title={Transformers: State-of-the-art natural language processing},
  author={Wolf, Thomas and Debut, Lysandre and Sanh, Victor and Chaumond, Julien and Delangue, Clement and Moi, Anthony and Cistac, Pierric and Rault, Tim and Louf, R{\'e}mi and Funtowicz, Morgan and others},
  booktitle={Proceedings of the 2020 conference on empirical methods in natural language processing: system demonstrations},
  pages={38--45},
  year={2020}
}

@article{baevski2020wav2vec,
  title={wav2vec 2.0: A framework for self-supervised learning of speech representations},
  author={Baevski, Alexei and Zhou, Yuhao and Mohamed, Abdelrahman and Auli, Michael},
  journal={Advances in neural information processing systems},
  volume={33},
  pages={12449--12460},
  year={2020}
}

@article{hsu2021hubert,
  title={Hubert: Self-supervised speech representation learning by masked prediction of hidden units},
  author={Hsu, Wei-Ning and Bolte, Benjamin and Tsai, Yao-Hung Hubert and Lakhotia, Kushal and Salakhutdinov, Ruslan and Mohamed, Abdelrahman},
  journal={IEEE/ACM transactions on audio, speech, and language processing},
  volume={29},
  pages={3451--3460},
  year={2021},
  publisher={IEEE}
}

@article{zhang2021survey,
  title={A survey on multi-task learning},
  author={Zhang, Yu and Yang, Qiang},
  journal={IEEE transactions on knowledge and data engineering},
  volume={34},
  number={12},
  pages={5586--5609},
  year={2021},
  publisher={IEEE}
}

@inproceedings{collobert2008unified,
  title={A unified architecture for natural language processing: Deep neural networks with multitask learning},
  author={Collobert, Ronan and Weston, Jason},
  booktitle={Proceedings of the 25th international conference on Machine learning},
  pages={160--167},
  year={2008}
}

@article{zhang2017cross,
  title={Cross-corpus acoustic emotion recognition with multi-task learning: Seeking common ground while preserving differences},
  author={Zhang, Biqiao and Provost, Emily Mower and Essl, Georg},
  journal={IEEE Transactions on Affective Computing},
  volume={10},
  number={1},
  pages={85--99},
  year={2017},
  publisher={IEEE}
}

@article{sener2018multi,
  title={Multi-task learning as multi-objective optimization},
  author={Sener, Ozan and Koltun, Vladlen},
  journal={Advances in neural information processing systems},
  volume={31},
  year={2018}
}

@inproceedings{standley2020tasks,
  title={Which tasks should be learned together in multi-task learning?},
  author={Standley, Trevor and Zamir, Amir and Chen, Dawn and Guibas, Leonidas and Malik, Jitendra and Savarese, Silvio},
  booktitle={International conference on machine learning},
  pages={9120--9132},
  year={2020},
  organization={PMLR}
}

@article{latif2022multitask,
  title={Multitask learning from augmented auxiliary data for improving speech emotion recognition},
  author={Latif, Siddique and Rana, Rajib and Khalifa, Sara and Jurdak, Raja and Schuller, Bj{\"o}rn W},
  journal={IEEE Transactions on Affective Computing},
  volume={14},
  number={4},
  pages={3164--3176},
  year={2022},
  publisher={IEEE}
}

@inproceedings{cai2021speech,
  title={Speech emotion recognition with multi-task learning.},
  author={Cai, Xingyu and Yuan, Jiahong and Zheng, Renjie and Huang, Liang and Church, Kenneth},
  booktitle={Interspeech},
  volume={2021},
  pages={4508--4512},
  year={2021},
  organization={Brno}
}

@article{busso2008iemocap,
	title={{IEMOCAP}: Interactive emotional dyadic motion capture database},
	author={Busso, Carlos and Bulut, Murtaza and Lee, Chi-Chun and Kazemzadeh, Abe and Mower, Emily and Kim, Samuel and Chang, Jeannette N and Lee, Sungbok and Narayanan, Shrikanth S},
	journal={Language resources and evaluation},
	volume={42},
	number={4},
	pages={335--359},
	year={2008},
	publisher={Springer}
}

@inproceedings{gao2024enhancing,
  title={Enhancing Two-Stage Finetuning for Speech Emotion Recognition Using Adapters},
  author={Yuan Gao and Hao Shi and Chenhui Chu and Tatsuya Kawahara},
  booktitle={ICASSP 2024-2024 IEEE International Conference on Acoustics, Speech and Signal Processing (ICASSP)},
  pages={11316--11320},
  year={2024},
  organization={IEEE}
}

@inproceedings{triantafyllopoulos2021deep,
  title={Deep speaker conditioning for speech emotion recognition},
  author={Triantafyllopoulos, Andreas and Liu, Shuo and Schuller, Bj{\"o}rn W},
  booktitle={2021 IEEE international conference on multimedia and expo (ICME)},
  pages={1--6},
  year={2021},
  organization={IEEE}
}

@article{zhang2019persemon,
  title={PersEmoN: A deep network for joint analysis of apparent personality, emotion and their relationship},
  author={Zhang, Le and Peng, Songyou and Winkler, Stefan},
  journal={IEEE Transactions on Affective Computing},
  volume={13},
  number={1},
  pages={298--305},
  year={2019},
  publisher={IEEE}
}

@article{gosling2003very,
  title={A very brief measure of the Big-Five personality domains},
  author={Gosling, Samuel D and Rentfrow, Peter J and Swann Jr, William B},
  journal={Journal of Research in personality},
  volume={37},
  number={6},
  pages={504--528},
  year={2003},
  publisher={Elsevier}
}

@book{fleiss2013statistical,
  title={Statistical methods for rates and proportions},
  author={Fleiss, Joseph L and Levin, Bruce and Paik, Myunghee Cho},
  year={2013},
  publisher={john wiley \& sons}
}

@article{cohen2009pearson,
  title={Pearson correlation coefficient},
  author={Cohen, Israel and Huang, Yiteng and Chen, Jingdong and Benesty, Jacob and Benesty, Jacob and Chen, Jingdong and Huang, Yiteng and Cohen, Israel},
  journal={Noise reduction in speech processing},
  pages={1--4},
  year={2009},
  publisher={Springer}
}

@article{ramakrishnan2013speech,
  title={Speech emotion recognition approaches in human computer interaction},
  author={Ramakrishnan, Srinivasan and El Emary, Ibrahiem MM},
  journal={Telecommunication Systems},
  volume={52},
  pages={1467--1478},
  year={2013},
  publisher={Springer}
}

@article{chatterjee2021real,
  title={Real-time speech emotion analysis for smart home assistants},
  author={Chatterjee, Rajdeep and Mazumdar, Saptarshi and Sherratt, R Simon and Halder, Rohit and Maitra, Tanmoy and Giri, Debasis},
  journal={IEEE Transactions on Consumer Electronics},
  volume={67},
  number={1},
  pages={68--76},
  year={2021},
  publisher={IEEE}
}

@inproceedings{dhuheir2021emotion,
  title={Emotion recognition for healthcare surveillance systems using neural networks: A survey},
  author={Dhuheir, Marwan and Albaseer, Abdullatif and Baccour, Emna and Erbad, Aiman and Abdallah, Mohamed and Hamdi, Mounir},
  booktitle={2021 International Wireless Communications and Mobile Computing (IWCMC)},
  pages={681--687},
  year={2021},
  organization={IEEE}
}

@article{costa1999five,
  title={A five-factor theory of personality},
  author={Costa, PT and McCrae, RR},
  journal={Handbook of personality: Theory and research},
  volume={2},
  number={01},
  pages={1999},
  year={1999},
  publisher={Guildford Press New York}
}

@article{simha2020big,
  title={The big 5 personality traits and willingness to justify unethical behavior—a cross-national examination},
  author={Simha, Aditya and Parboteeah, K Praveen},
  journal={Journal of Business Ethics},
  volume={167},
  pages={451--471},
  year={2020},
  publisher={Springer}
}

@article{deniz2011investigation,
  title={An investigation of decision making styles and the five-factor personality traits with respect to attachment styles.},
  author={Deniz, Mehmet},
  journal={Educational Sciences: Theory and Practice},
  volume={11},
  number={1},
  pages={105--113},
  year={2011},
  publisher={ERIC}
}

@article{curtis2015relationship,
  title={The relationship between Big-5 personality traits and cognitive ability in older adults--a review},
  author={Curtis, Rachel G and Windsor, Tim D and Soubelet, Andrea},
  journal={Aging, Neuropsychology, and Cognition},
  volume={22},
  number={1},
  pages={42--71},
  year={2015},
  publisher={Taylor \& Francis}
}

@article{li2023transfer,
  title={Transfer Learning for Personality Perception via Speech Emotion Recognition},
  author={Li, Yuanchao and Bell, Peter and Lai, Catherine},
  journal={arXiv preprint arXiv:2305.16076},
  year={2023}
}

@article{cattell1992handbook,
  title={Handbook for the sixteen personality factor questionnaire (16 PF)},
  author={Cattell, Raymond Bernard and Eber, Herbert W and Tatsuoka, Maurice M},
  journal={(No Title)},
  year={1992}
}

@article{ashton2007empirical,
  title={Empirical, theoretical, and practical advantages of the HEXACO model of personality structure},
  author={Ashton, Michael C and Lee, Kibeom},
  journal={Personality and social psychology review},
  volume={11},
  number={2},
  pages={150--166},
  year={2007},
  publisher={SAGE publications Sage CA: Los Angeles, CA}
}

@article{azucar2018predicting,
  title={Predicting the Big 5 personality traits from digital footprints on social media: A meta-analysis},
  author={Azucar, Danny and Marengo, Davide and Settanni, Michele},
  journal={Personality and individual differences},
  volume={124},
  pages={150--159},
  year={2018},
  publisher={Elsevier}
}

@book{myers2010gifts,
  title={Gifts differing: Understanding personality type},
  author={Myers, Isabel Briggs and Myers, Peter B},
  year={2010},
  publisher={Nicholas Brealey}
}

@article{costa2008revised,
  title={The revised neo personality inventory (neo-pi-r)},
  author={Costa, Paul T and McCrae, Robert R},
  journal={The SAGE handbook of personality theory and assessment},
  volume={2},
  number={2},
  pages={179--198},
  year={2008}
}

@book{vernon2006thinking,
  title={Thinking, feeling, behaving: an emotional education curriculum for children. Grades 1-6},
  author={Vernon, Ann},
  year={2006},
  publisher={Research Press}
}

@article{stemmler2010personality,
  title={Personality, emotion, and individual differences in physiological responses},
  author={Stemmler, Gerhard and Wacker, Jan},
  journal={Biological psychology},
  volume={84},
  number={3},
  pages={541--551},
  year={2010},
  publisher={Elsevier}
}

@inproceedings{wu2017implicit,
  title={Implicit acquisition of user personality for augmenting recommender systems},
  author={Wu, Wen},
  booktitle={Companion Proceedings of the 22nd International Conference on Intelligent User Interfaces},
  pages={201--204},
  year={2017}
}

@inproceedings{kahn2020libri,
  title={Libri-light: A benchmark for asr with limited or no supervision},
  author={Kahn, Jacob and Riviere, Morgane and Zheng, Weiyi and Kharitonov, Evgeny and Xu, Qiantong and Mazar{\'e}, Pierre-Emmanuel and Karadayi, Julien and Liptchinsky, Vitaliy and Collobert, Ronan and Fuegen, Christian and others},
  booktitle={ICASSP 2020-2020 IEEE International Conference on Acoustics, Speech and Signal Processing (ICASSP)},
  pages={7669--7673},
  year={2020},
  organization={IEEE}
}

@article{langston1997beliefs,
  title={Beliefs and the Big Five: Cognitive bases of broad individual differences in personality},
  author={Langston, Christopher A and Sykes, W Eric},
  journal={Journal of Research in Personality},
  volume={31},
  number={2},
  pages={141--165},
  year={1997},
  publisher={Elsevier}
}

@article{vaswani2017attention,
  title={Attention is all you need},
  author={Vaswani, Ashish and Shazeer, Noam and Parmar, Niki and Uszkoreit, Jakob and Jones, Llion and Gomez, Aidan N and Kaiser, {\L}ukasz and Polosukhin, Illia},
  journal={Advances in neural information processing systems},
  volume={30},
  year={2017}
}

@article{guidi2019analysis,
  title={Analysis of speech features and personality traits},
  author={Guidi, Andrea and Gentili, Claudio and Scilingo, Enzo Pasquale and Vanello, Nicola},
  journal={Biomedical signal processing and control},
  volume={51},
  pages={1--7},
  year={2019},
  publisher={Elsevier}
}

@inproceedings{sagha2017effect,
  title={The effect of personality trait, age, and gender on the performance of automatic speech valence recognition},
  author={Sagha, Hesam and Deng, Jun and Schuller, Bj{\"o}rn},
  booktitle={2017 seventh international conference on affective computing and intelligent interaction (ACII)},
  pages={86--91},
  year={2017},
  organization={IEEE}
}

@inproceedings{graves2006connectionist,
  title={Connectionist temporal classification: labelling unsegmented sequence data with recurrent neural networks},
  author={Graves, Alex and Fern{\'a}ndez, Santiago and Gomez, Faustino and Schmidhuber, J{\"u}rgen},
  booktitle={Proceedings of the 23rd international conference on Machine learning},
  pages={369--376},
  year={2006}
}

@incollection{fisher1970statistical,
  title={Statistical methods for research workers},
  author={Fisher, Ronald Aylmer},
  booktitle={Breakthroughs in statistics: Methodology and distribution},
  pages={66--70},
  year={1970},
  publisher={Springer}
}

@article{khan2025memocmt,
  title={MemoCMT: multimodal emotion recognition using cross-modal transformer-based feature fusion},
  author={Khan, Mustaqeem and Tran, Phuong-Nam and Pham, Nhat Truong and El Saddik, Abdulmotaleb and Othmani, Alice},
  journal={Scientific reports},
  volume={15},
  number={1},
  pages={5473},
  year={2025},
  publisher={Nature Publishing Group UK London}
}

@article{khan2025joint,
  title={Joint Multi-Scale Multimodal Transformer for Emotion Using Consumer Devices},
  author={Khan, Mustaqeem and Ahmad, Jamil and Gueaieb, Wail and De Masi, Giulia and Karray, Fakhri and El Saddik, Abdulmotaleb},
  journal={IEEE Transactions on Consumer Electronics},
  year={2025},
  publisher={IEEE}
}

@inproceedings{messaoudi2024modeling,
  title={Modeling continuous emotions in text data using IEMOCAP database},
  author={Messaoudi, Awatef and Boughrara, Hayet and Lachiri, Zied},
  booktitle={2024 IEEE 7th International Conference on Advanced Technologies, Signal and Image Processing (ATSIP)},
  volume={1},
  pages={397--402},
  year={2024},
  organization={IEEE}
}

@inproceedings{zhou2024learning,
  title={Learning arousal-valence representation from categorical emotion labels of speech},
  author={Zhou, Enting and Zhang, You and Duan, Zhiyao},
  booktitle={ICASSP 2024-2024 IEEE International Conference on Acoustics, Speech and Signal Processing (ICASSP)},
  pages={12126--12130},
  year={2024},
  organization={IEEE}
}

@inproceedings{vlasenko2024comparing,
  title={Comparing data-driven and handcrafted features for dimensional emotion recognition},
  author={Vlasenko, Bogdan and Vyas, Sargam and Doss, Mathew Magimai-},
  booktitle={ICASSP 2024-2024 IEEE International Conference on Acoustics, Speech and Signal Processing (ICASSP)},
  pages={11841--11845},
  year={2024},
  organization={IEEE}
}

@article{wagner2023dawn,
  title={Dawn of the transformer era in speech emotion recognition: closing the valence gap},
  author={Wagner, Johannes and Triantafyllopoulos, Andreas and Wierstorf, Hagen and Schmitt, Maximilian and Burkhardt, Felix and Eyben, Florian and Schuller, Bj{\"o}rn W},
  journal={IEEE Transactions on Pattern Analysis and Machine Intelligence},
  volume={45},
  number={9},
  pages={10745--10759},
  year={2023},
  publisher={IEEE}
}

@inproceedings{ispas2023multi,
  title={A multi-task, multi-modal approach for predicting categorical and dimensional emotions},
  author={Ispas, Alex-R{\u{a}}zvan and Deschamps-Berger, Th{\'e}o and Devillers, Laurence},
  booktitle={Companion Publication of the 25th International Conference on Multimodal Interaction},
  pages={311--317},
  year={2023}
}

@inproceedings{srinivasan2022representation,
  title={Representation learning through cross-modal conditional teacher-student training for speech emotion recognition},
  author={Srinivasan, Sundararajan and Huang, Zhaocheng and Kirchhoff, Katrin},
  booktitle={ICASSP 2022-2022 IEEE International Conference on Acoustics, Speech and Signal Processing (ICASSP)},
  pages={6442--6446},
  year={2022},
  organization={IEEE}
}

@article{lawrence1989concordance,
  title={A concordance correlation coefficient to evaluate reproducibility},
  author={Lawrence, I and Lin, Kuei},
  journal={Biometrics},
  pages={255--268},
  year={1989},
  publisher={JSTOR}
}

\begin{IEEEbiography}[{\includegraphics[width=1in,height=1.25in,clip,keepaspectratio]{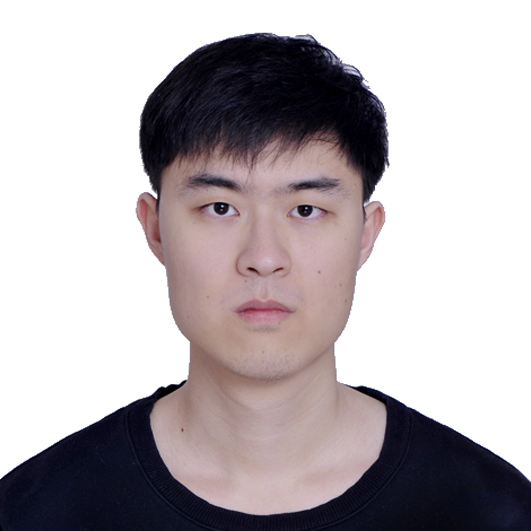}}]{Dr. Yuan Gao} received the B.E. degree in school of computer science from Hebei University of Technology, Tianjin, China, in 2019, and the M.S. degree from both Tianjin University, Tianjin, China, and the Japan Advanced Institute of Science and Technology, Ishikawa, Japan, in 2022. He received the Ph.D. degree in Intelligence Science and Technology from Kyoto University, Kyoto, Japan, in 2025. He is currently a Research Scientist with SB Intuitions. His research interests include speech signal processing and multimodal emotion recognition.
\end{IEEEbiography}

\begin{IEEEbiography}[{\includegraphics[width=1in,height=1.25in,clip,keepaspectratio]{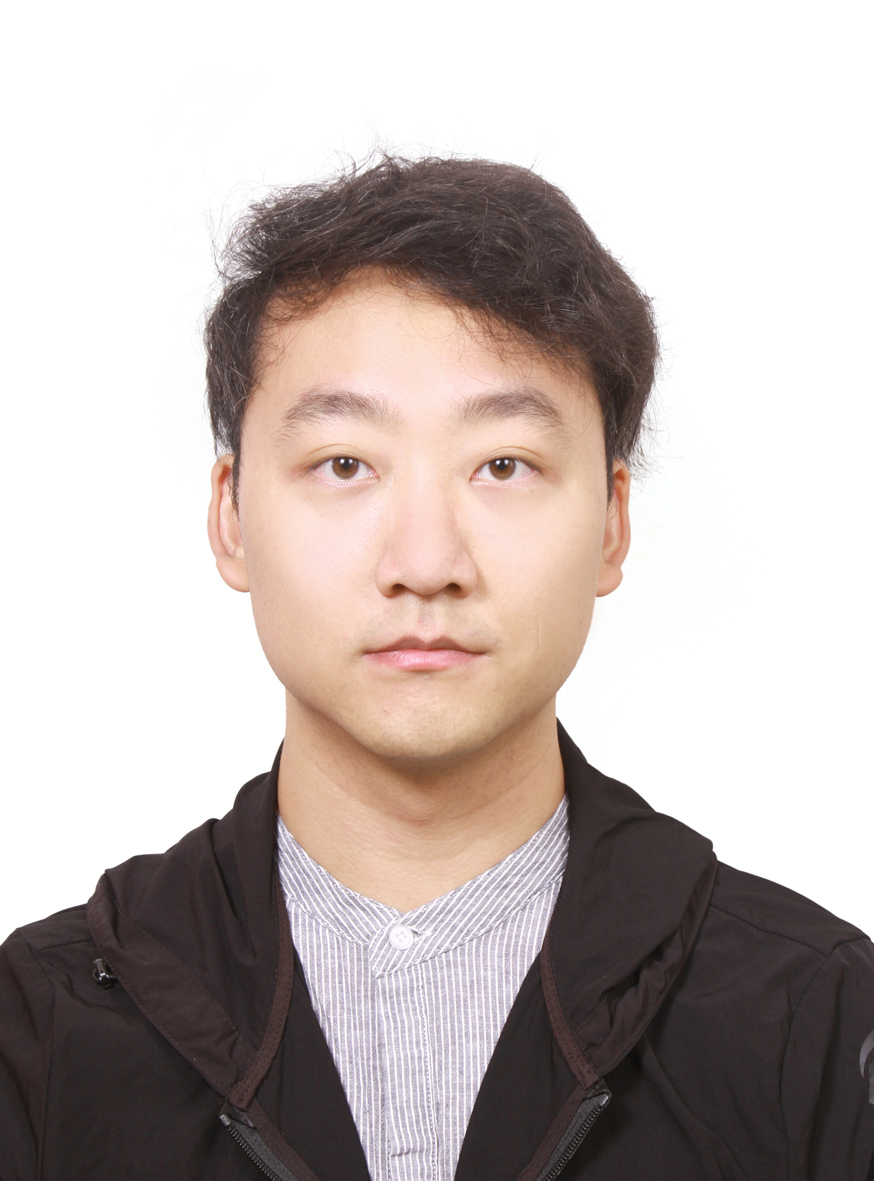}}]{Dr. Hao Shi} (Member, IEEE) received the B.E. degree in computer science from Southwest Jiaotong University, Chengdu, China, in 2018, and the M.S. degree in computer science from Tianjin University, Tianjin, China, in 2021. He received the Ph.D. degree in informatics from Kyoto University, Kyoto, Japan, in 2024. From October 2024 to March 2025, he was a researcher at Kyoto University. He is currently a research scientist with SB Intuitions. His research interests include automatic speech recognition and speech enhancement.
\end{IEEEbiography}

\begin{IEEEbiography}[{\includegraphics[width=1in,height=1.25in,clip,keepaspectratio]{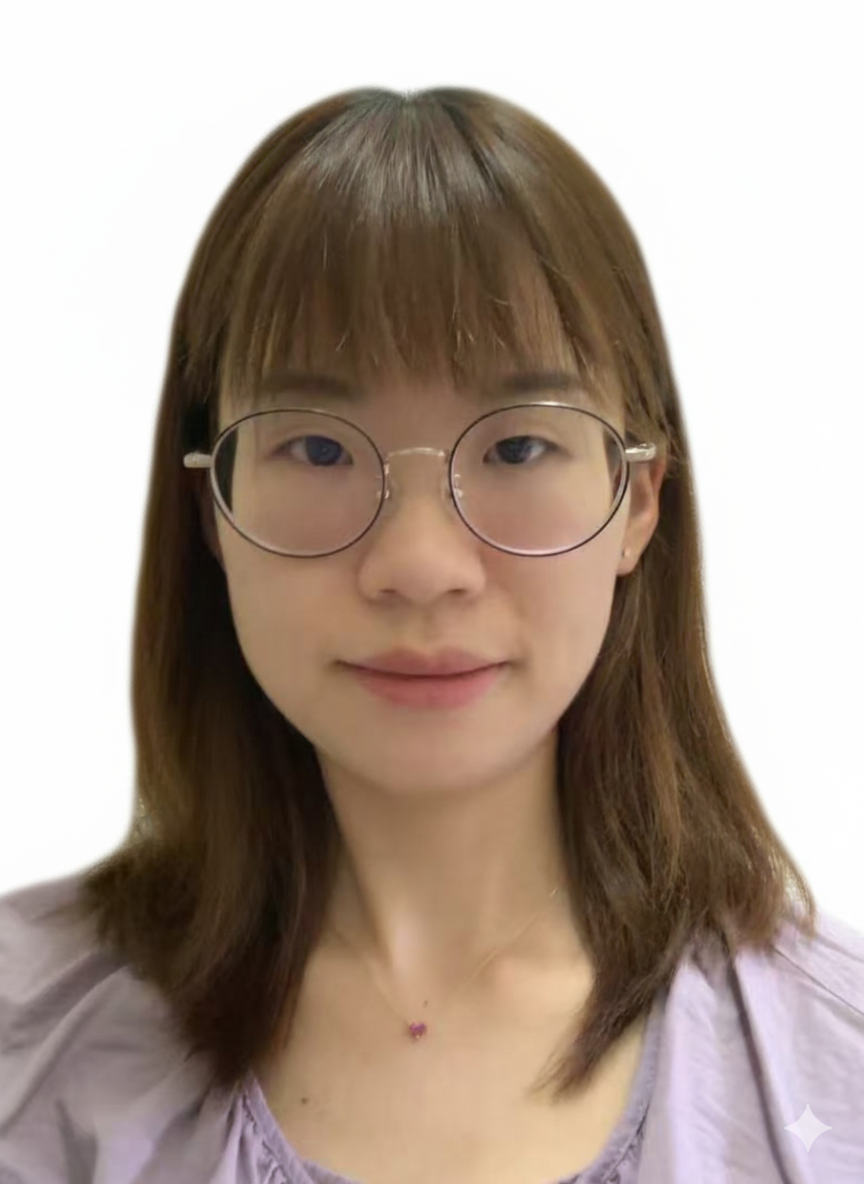}}]{Dr. Yahui Fu} received her M.S. degrees from both Tianjin University, Tianjin, China, and the Japan Advanced Institute of Science and Technology, Ishikawa, Japan, in 2021. She received her Ph.D. degree in Intelligence Science and Technology from Kyoto University, Kyoto, Japan, in 2024. She is currently a postdoctoral researcher at Kyoto University. Her research interests include dialogue systems and multimodal emotion recognition.
\end{IEEEbiography}

\begin{IEEEbiography}[{\includegraphics[width=1in,height=1.25in,clip,keepaspectratio]{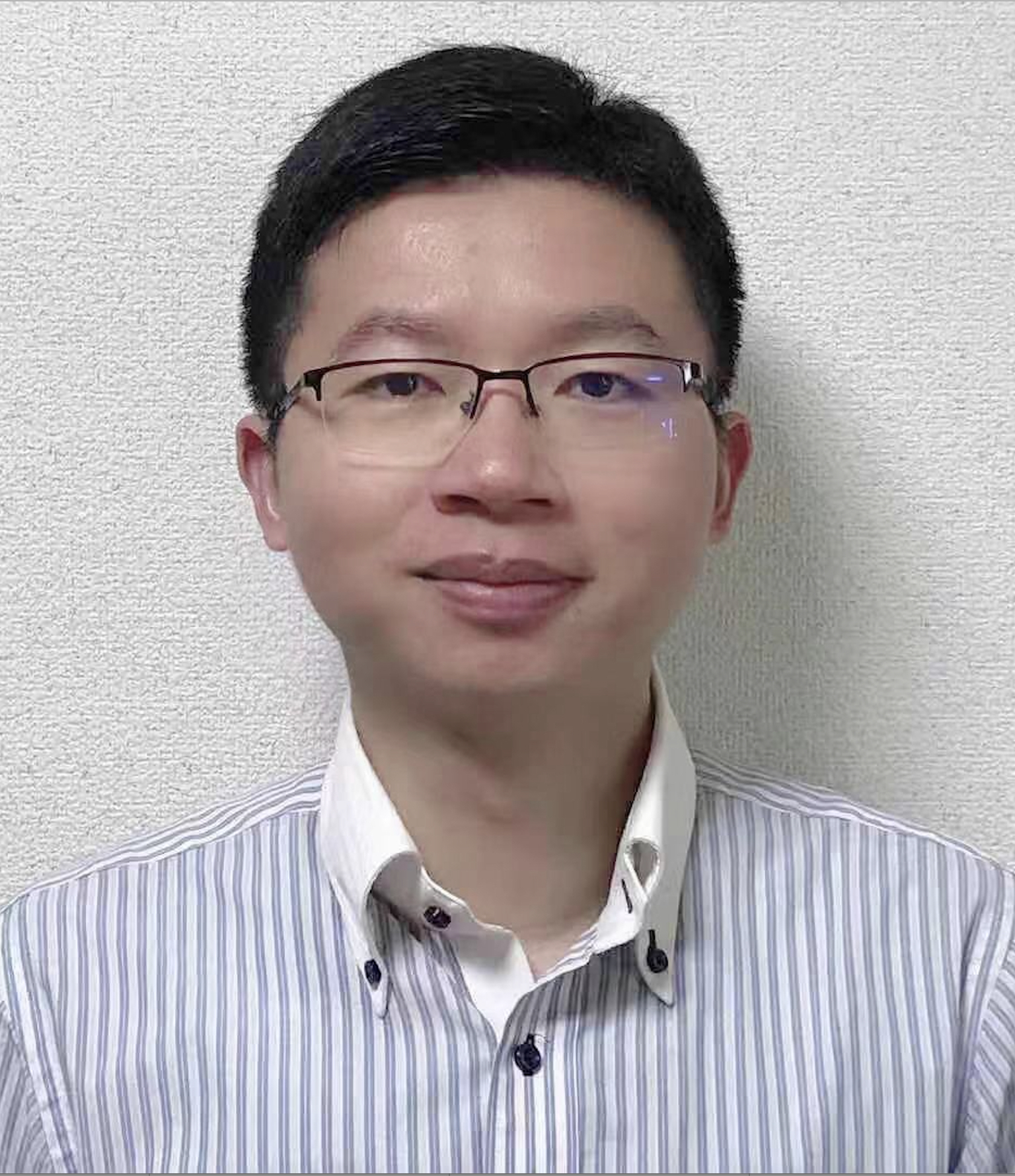}}]{Chenhui Chu} received his B.S. in software engineering from 
Chongqing University in 2008, and his M.S. and Ph.D. in Informatics 
from Kyoto University in 2012 and 2015, respectively. He is currently 
a associate professor at Kyoto University. His 
research interests include natural language processing, particularly 
machine translation and multimodal machine learning. 
\end{IEEEbiography}

\begin{IEEEbiography}[{\includegraphics[width=1in,height=1.25in,clip,keepaspectratio]{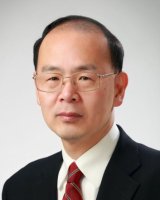}}]{Tatsuya Kawahara}  (Fellow, IEEE) received B.E. in 1987, M.E. in 1989, and Ph.D. in 1995, all in information science, from Kyoto University, Kyoto, Japan.
From 1995 to 1996, he was a Visiting Researcher at Bell Laboratories, Murray Hill, NJ, USA. Currently, he is a Professor of School of Informatics, Kyoto University. From 2020 to 2023, he was the Dean of the School. Before that, he was also an Invited Researcher at ATR and NICT.
He has published more than 450 academic papers on automatic speech recognition, spoken language processing, and spoken dialogue systems. He has been conducting several projects including open-source speech recognition software Julius, the automatic transcription system deployed in the Japanese Parliament (Diet), and the autonomous android ERICA.

Dr. Kawahara received the Commendation for Science and Technology by the Minister of Education, Culture, Sports, Science and Technology (MEXT) in 2012.
From 2003 to 2006, he was a member of IEEE SPS Speech Technical Committee. He was a General Chair of IEEE ASRU 2007 and is a General Chair of SIGdial 2024. He also served as a Tutorial Chair of INTERSPEECH 2010, a Local Arrangement Chair of ICASSP 2012, and a General Chair of APSIPA ASC 2020.
He was an editorial board member of Elsevier Journal of Computer Speech and Language and IEEE/ACM Transactions on Audio, Speech, and Language Processing. From 2018 to 2021, he was the Editor-in-Chief of APSIPA Transactions on Signal and Information Processing.
Dr. Kawahara is the President of APSIPA, the Secretary General of ISCA, and a Fellow of IEEE.
\end{IEEEbiography}

\end{document}